\documentclass{article}

\usepackage{arxiv}

\usepackage[utf8]{inputenc} 
\usepackage[T1]{fontenc}    
\usepackage{hyperref}       
\usepackage{url}            
\usepackage{booktabs}       
\usepackage{amsfonts}       
\usepackage{nicefrac}       
\usepackage{microtype}      
\usepackage{lipsum}
\usepackage{graphicx}
\usepackage{multirow}
\usepackage{amsmath}
\graphicspath{ {./images/} }

\title{Cross-Modal Epileptic Signal Harmonization: Frequency Domain Mapping Quantization for Pre-training a Unified Neurophysiological Transformer}

\author{
 Runkai Zhang \\
  Key Laboratory of Child Development and Learning Science of Ministry of Education\\
  School of Biological Science \& Medical Engineering \\
  Southeast University\\
  Nanjing, 210096 \\
  Jiangsu, PR China \\
   \And
 Hua Yu \\
  Department of Cardiology, The First Afﬁliated Hospital of USTC\\
  Division of Life Sciences and Medicine\\
  University of Science and Technology of China\\
  Hefei, 230001 \\
  Jiangsu, PR China \\
    \And
 John Q. Gan \\
  School of Computer Science and Electronic Engineering\\
  University of Essex\\
  Colchester CO4 3SQ \\
  UK\\
    \AND
 Haixian Wang* \\
  Key Laboratory of Child Development and Learning Science of Ministry of Education\\
  School of Biological Science \& Medical Engineering \\
  Southeast University\\
  Nanjing, 210096 \\
  Jiangsu, PR China \\
  \texttt{hxwang@seu.edu.cn}
}

\begin{document}
\maketitle
\begin{abstract}
Scalp electroencephalography (EEG) and intracranial EEG (iEEG) are vital for epilepsy diagnosis and treatment. Their unified analysis offers the potential to harness the complementary strengths of each modality but is challenging due to variations in recording montages, amplitude and signal-to-noise ratio (SNR), and frequency components. To address the aforementioned challenges, this paper introduces EpiNT, a novel Transformer-based pre-trained model for unified EEG and iEEG analysis. EpiNT employs channel-independent modeling with masked autoencoders (MAE) and vector quantization (VQ), along with a frequency domain mapping quantizer to capture crucial frequency features. Pre-trained on over 2,700 hours of multi-modal clinical neurophysiological data from 1,199 patients, EpiNT outperformed both randomly initialized models and other pre-trained methods on six downstream classification tasks, demonstrating robust representation learning capabilities. This work presents a promising approach for unified epilepsy neurophysiology analysis.
\end{abstract}


\section{Introduction}
Epilepsy is a chronic neurological disorder characterized by  recurrent, unprovoked seizures \cite{Rana2018inflammation}. As one of the most common neurological disorders, it affects approximately 70 million people worldwide \cite{thijs2019epilepsy}, posing significant clinical and social challenges \cite{beghi2020epidemiology}. Understanding and managing epilepsy relies heavily on the ability to directly assess brain electrical activity \cite{maturana2020critical}.  Neurophysiological signals, recorded by techniques like electroencephalography (EEG) \cite{abou2022noninvasive}, electrocorticography (ECoG) and stereo-electroencephalography (SEEG) \cite{gunnarsdottir2022source}, provides invaluable insights into this neurological disease.

Recent advances in deep learning (DL) have revolutionized the analysis of such neurophysiological signals, particularly in epilepsy \cite{roy2019deep}. DL models have achieved impressive performance in key clinical tasks such as seizure prediction \cite{shi2023b2}, seizure detection \cite{li2020epileptic}, interictal epileptiform discharges (IEDs) identification \cite{lin2024vepinet} and high-frequency oscillations (HFOs) identification \cite{zhang2024pyhfo}. Furthermore, DL models have also been used to predict treatment outcomes of antiepileptic drugs and identify relevant biomarkers from EEG data \cite{zhang2024oxcarnet}. However, these models are typically designed for specific tasks, limiting their ability to generalize across diverse clinical applications and patient populations.

To address this limitation, pre-training has emerged as a powerful paradigm in machine learning, enabling models to learn generalizable representations from large-scale unlabeled data. Widely adopted in computer vision \cite{he2016deep}, natural language processing \cite{devlin2018bert}, and neural signal analysis \cite{banville2019self}, pre-training improves performance in downstream tasks. Existing EEG pre-training methods largely follow two paradigms: direct reconstruction and vector quantization (VQ)-based modeling. Reconstruction-based methods aim to recover masked segments of raw \cite{cai2024mae} or spectral \cite{wang2023brainbert} signals. However, the noise and redundancy in neural data often limit the effectiveness of direct reconstruction. In contrast, VQ-based methods learn compact, discrete representations that are less sensitive to noise and more stable during training. Jiang et al. \cite{jiang2024large} applied VQ to tokenize EEG into discrete sequences, while Gui et al. \cite{gui2024vector} used random projection for efficient quantization. pre-training modeling strategies also vary by signal modality: scalp EEG studies (e.g., MMM \cite{yi2024learning}) emphasize montage-invariant learning, whereas intracranial EEG (iEEG) approaches (e.g., Brant \cite{zhang2024brant}) focus on spectral encoding to capture richer high-frequency features. Despite these differences, all methods share the goal of learning robust and generalizable neural representations \cite{kim2024towards}.

\begin{figure}[h] 
    \centering
    \includegraphics[width=0.9\textwidth]{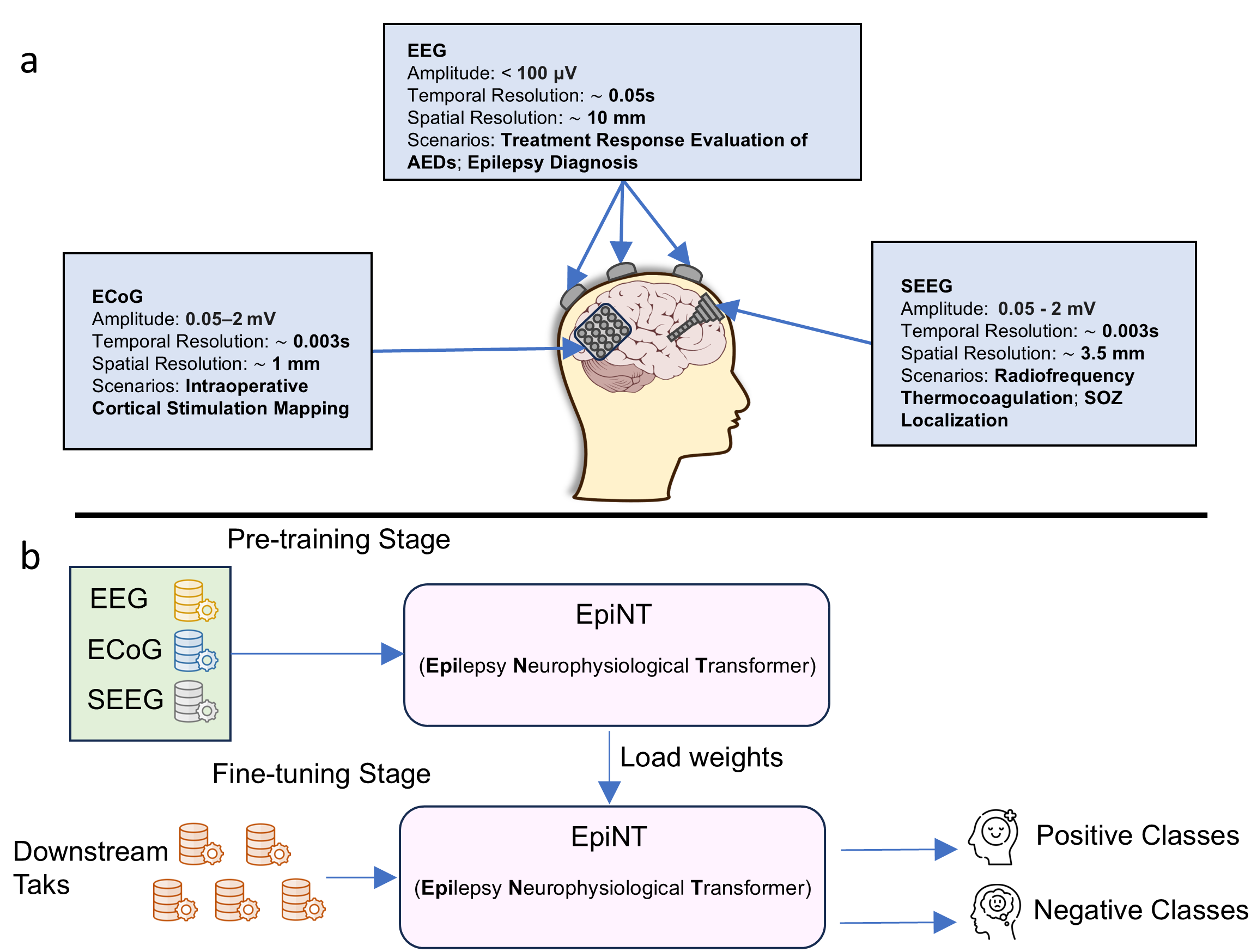}
    \caption{\textbf{Characteristics of different electrophysiological modalities and the proposed pre-trained model.} (a) Scalp EEG, electrocorticography (ECoG), and stereo-electroencephalography (SEEG) differ in signal amplitude, temporal and spatial resolution, and typical clinical applications. (b) The proposed EpiNT leverages EEG, ECoG, and SEEG to pre-train, and fine-tuned using downstream taks.}
    \label{fig:intro} 
\end{figure}

Nonetheless, pre-training models specifically designed for the neurophysiological characteristics of epilepsy are still limited. Clinically relevant epileptic biomarkers are often sparse and localized, manifesting only in a few EEG channels during focal seizures or in isolated events such as IEDs and HFOs. This sparsity exacerbates class imbalance in downstream tasks. Furthermore, significant discrepancies between scalp and intracranial recordings complicate unified modeling (Figure~\ref{fig:intro}a). These include:
\begin{itemize}
    \item Montage variability: Scalp EEG typically employs standardized electrode placements based on the 10-20 system or its derivatives \cite{hatlestad2023reliable}, while iEEG involves patient-specific electrode placements \cite{mitsuhashi2020effects}.
    \item Signal amplitude and SNR: Scalp EEG has low amplitude (10–100 µV) and is prone to artifacts \cite{saha2021progress}, whereas iEEG has higher amplitude (hundreds of µV to several mV) and better SNR due to proximity to neural sources \cite{mercier2022advances}.
    \item Frequency content variations: Volume conduction limits scalp EEG's sensitivity to higher frequencies \cite{shakeshaft2022heterogeneity}, while iEEG captures a broader spectrum, including clinically relevant HFOs such as ripples (80-250 Hz) and fast ripples (250-500 Hz) \cite{sakakura2023developmental}.
\end{itemize}

In this study, we introduce EpiNT (\textbf{Epi}lepsy \textbf{N}europhysiological \textbf{T}ransformer), a unified pre-trained Transformer model tailored for epilepsy-related EEG and iEEG analysis (Figure \ref{fig:intro}b). EpiNT aims to address the lack of generalizable models that can robustly interpret both scalp and intracranial signals across diverse acquisition settings. Our design incorporates:
\begin{itemize}
    \item Channel-independent modeling to accommodate montage variability;
    \item  A hybrid pre-training strategy that combines masked autoencoding (MAE) and VQ to enhance robustness to amplitude and SNR differences;
    \item A novel frequency-domain quantizer to capture discrete spectral representations and address frequency content variability.
\end{itemize}
EpiNT is pre-trained on a curated corpus of over 2700 hours of EEG/iEEG recordings from 1199 patients. The unsupervised pre-training task involves predicting discrete representations of masked input signals to learn robust, modality-invariant features. For fine-tuning, the model is adapted using cross-entropy loss with either full or partial parameter updates. We validate EpiNT across five clinically relevant binary classification downstream tasks: seizure onset zone (SOZ) localization, seizure prediction, interictal epileptiform discharge (IED) detection, pathology classification, and seizure detection. Each downstream task involves determining whether a given input signal contains specific types of epileptiform activity relevant to clinical diagnosis and treatment planning. EpiNT demonstrates strong generalization in cross-subject settings and offers a unified, scalable foundation for automated epilepsy EEG interpretation, potentially streamlining clinical workflows across heterogeneous patient populations.

\section{Results}

\subsection{Datasets for Pre-training and Downstream Evaluation}\label{Dataset Characteristics}
Data is vital for effective model pre-training. Prior to reporting the performance of EpiNT, we describe the datasets used for both unsupervised pre-training and downstream evaluation. 

\begin{figure}[h]
    \centering
    \includegraphics[width=0.9\textwidth]{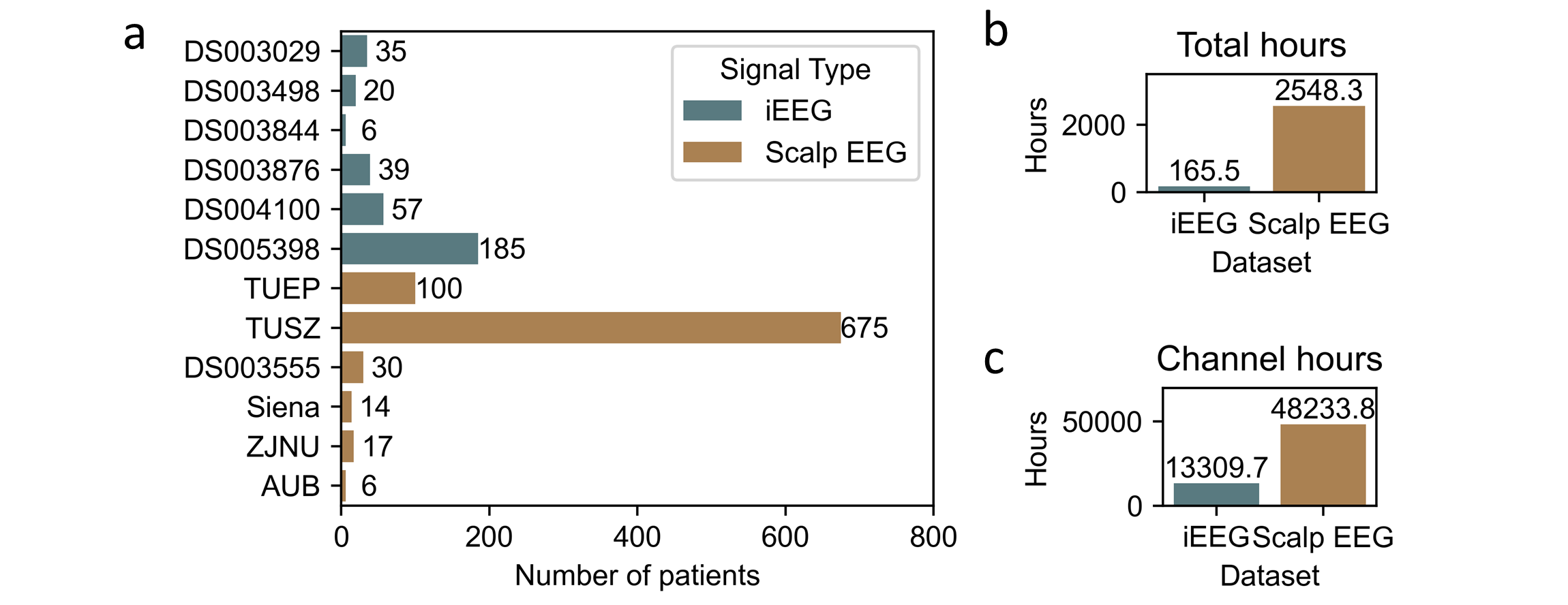}
    \caption{\textbf{Characteristics of pre-training dataset.} (a) Patient distribution reveals a substantial proportion of scalp EEG recordings compared to iEEG. (b) Total recording duration demonstrates a significantly higher volume of scalp EEG data (2548.3 hours) relative to iEEG (165.5 hours). (c) Upon partitioning into single-channel data, the disparity in data volume persists, with scalp EEG yielding 48233.8 hours and iEEG 13309.7 hours.}
    \label{fig:pre-train dataset characteristics} 
\end{figure}

We curated a large-scale dataset comprising 2,741.1 hours of EEG and iEEG recordings from 1,199 patients for unsupervised pre-training. The characteristics of this dataset are summarized in Table~\ref{tab:pretrain ds}. Due to the easier accessibility of scalp EEG, the pre-training dataset includes more patients with EEG than iEEG recordings (Figure~\ref{fig:pre-train dataset characteristics}a). Notably, only patients with confirmed epilepsy were selected from the TUEP dataset. As shown in Figure~\ref{fig:pre-train dataset characteristics}b, the scalp EEG subset consists of 2,548.3 hours of multi-channel data, which expands to 48,233.8 hours after segmentation into single-channel epochs. The iEEG subset contributes 165.5 hours of multi-channel data, yielding 13,309.7 single-channel hours.

Following pre-training, EpiNT was fine-tuned using a limited number of epochs for task-specific adaptation. To evaluate its generalizability, we designed five clinically relevant binary classification tasks across six datasets, each targeting the identification of specific epileptiform patterns in EEG or iEEG segments. For example, the seizure detection task using the HUH dataset involves classifying segments as ictal or inter-ictal. Table~\ref{tab:downstream ds} summarizes all downstream datasets and task definitions. All evaluations used cross-subject splits, with training and test patient assignments detailed in Supplementary Table 1. Sample distributions for each task are shown in Supplementary Figure 1, revealing class imbalance, especially in CHB-MIT (65,960 positive vs. 9,741 negative samples) and HUH (445,914 vs. 120,792).

\subsection{Impact of Pre-training Data Modality and Volume}
To understand how the modality and amount of pre-training data affect EpiNT's performance, we conducted two sets of experiments. 

\begin{figure}[h]
    \centering
    \includegraphics[width=0.5\textwidth]{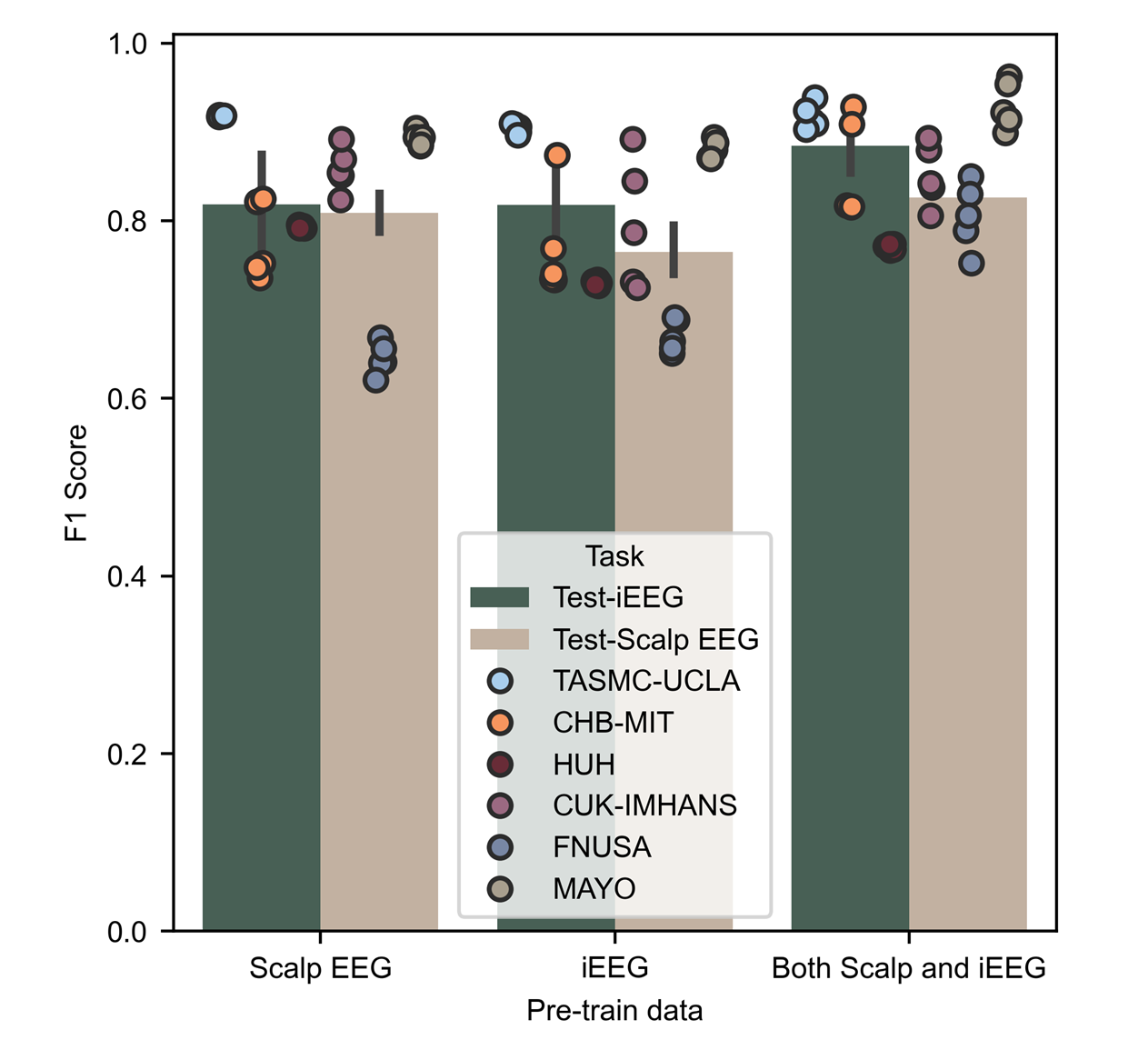} 
    \caption{\textbf{Performance evaluation of models pre-trained on different EEG modalities.} The intracranial-only model exhibits limited transferability to scalp EEG downstream tasks. While the scalp-only model shows strong generalization to scalp EEG tasks and comparable performance on iEEG tasks, but underperforms compared to the model pre-trained on a mixture of scalp and iEEG data. Pre-training on a mixed dataset of scalp and iEEG yields the highest overall performance across both downstream task modalities.}
    \label{fig:partial_eeg_ieeg} 
\end{figure}

First, we investigated the impact of pre-training data modality. We compared our primary model, which was pre-trained on a combination of scalp EEG and iEEG data, with two specialized models: one pre-trained only on scalp EEG data and another pre-trained only on iEEG data. We then assessed how well each pre-trained model performed on downstream tasks involving either scalp EEG or iEEG, aiming to see if the knowledge gained from one type of data could transfer to the other. As shown in Figure \ref{fig:partial_eeg_ieeg}, the model trained exclusively on iEEG data performed well on iEEG tasks but struggled more with scalp EEG tasks. The model pre-trained on a mixture of scalp and iEEG data consistently achieved the best performance, particularly on iEEG tasks. This indicates that combining both scalp and iEEG data during pre-training is crucial for optimal model performance and effective transfer of learned features across different types of brain activity recordings.

Second, we explored the effect of the amount of pre-training data. We varied the proportion of the total available pre-training data used for our EpiNT model (from 40\% to 100\%) and then evaluated its performance on various downstream tasks. Supplementary Figure 6 illustrates how the volume of the pre-training dataset influenced performance on these tasks. For tasks on the CHB-MIT and FNUSA datasets, increasing the amount of pre-training data consistently led to better performance, aligning with the common observation in deep learning that more data often yields superior results. However, tasks on the TASMC-UCLA and MAYO datasets exhibited a non-monotonic trend: performance initially decreased as more data was added, but then rebounded when the full dataset was utilized. This unexpected pattern suggests that sampling bias at intermediate data volumes, where smaller subsets may not accurately represent the data's full diversity, can lead to the learning of less robust features; conversely, the full dataset, by providing a more comprehensive sample, likely mitigates this bias and contributes to improved performance.

Our findings consistently demonstrate that pre-training a model on a diverse dataset is paramount for achieving optimal performance and robust cross-modality transfer. Furthermore, although increasing pre-training data generally improves performance, the benefits are not uniform across all tasks, underscoring the critical role of data quality in mitigating sampling biases and maximizing the utility of larger datasets.

\subsection{Interpretability and Behavior of EpiNT}
To understand how our pre-trained EpiNT model processes electrophysiological signals, we examined its internal layers and the characteristics of its learned features.

\begin{figure}[h] 
    \centering
    \includegraphics[width=0.9\textwidth]{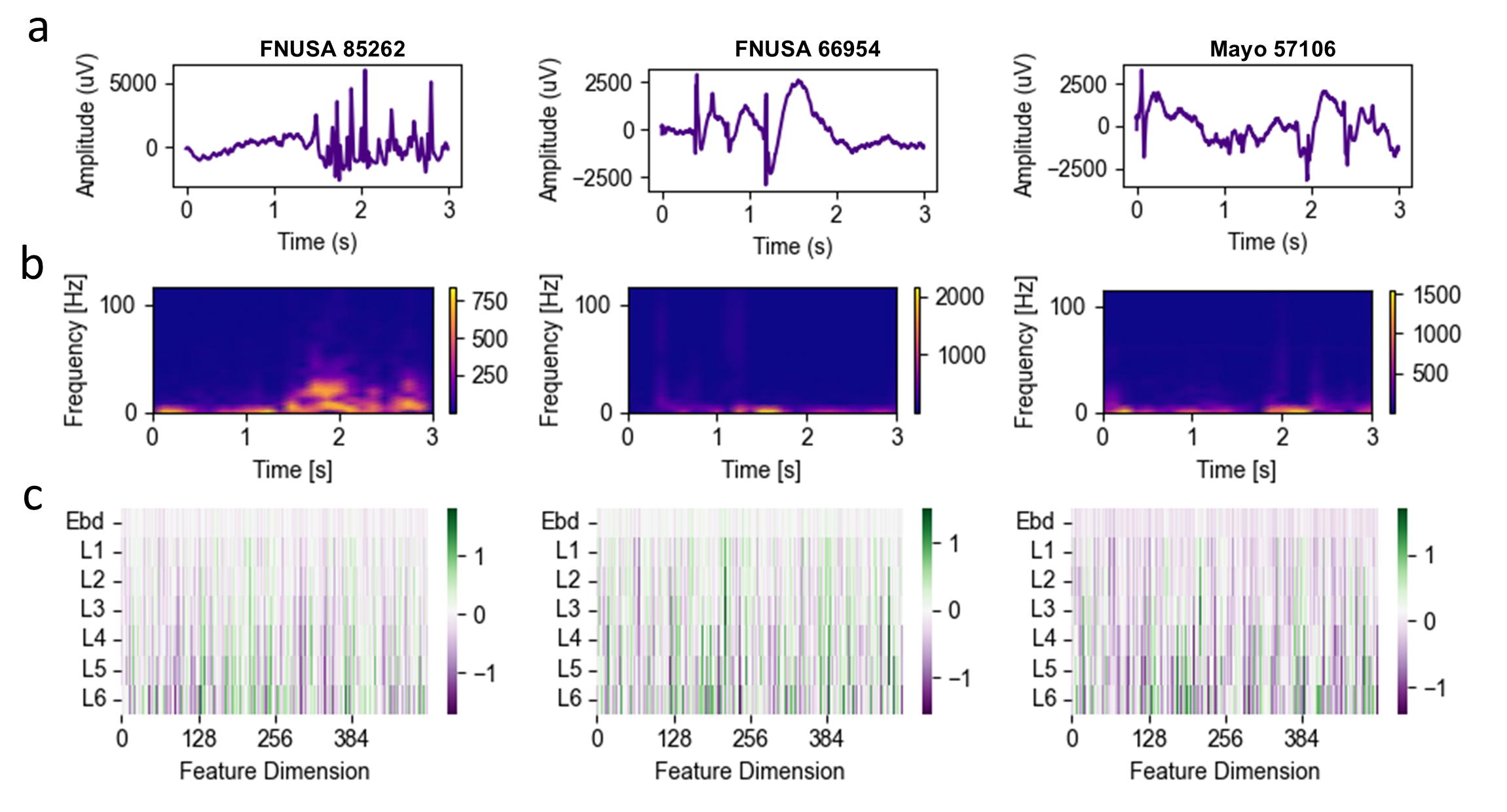} 
    \caption{\textbf{Visualization of representative electrophysiological signals and their corresponding feature representations across EpiNT layers.} (a) Raw waveforms of epileptic signals from the FNUSA and MAYO datasets, named by the indices in dataset. (b) STFT spectrograms of the raw waveforms. (c) Feature vector outputs from each layer of the EpiNT model. The initial embedding layer yielded feature vectors with magnitudes approaching zero across all dimensions, suggesting minimal initial feature differentiation. As activations propagated through the network, the Transformer layers exhibited feature vectors with increasingly polarized distributions, indicative of emerging abstract feature representations. Furthermore, a high degree of similarity in feature vectors between adjacent layers is noted, suggesting a gradual refinement of representations. Ebd: Embedding layer. L1-L6: Transformer Layer 1 to Layer 6.}
    \label{fig:examples}
\end{figure}

We first investigated how EpiNT transforms raw brain signals (e.g., IEDs and HFOs) as they pass through its layers. As seen in Figure \ref{fig:examples}, the initial layers generated very subtle feature representations. However, as the information moved deeper into the network, the Transformer layers progressively developed more distinct and abstract feature representations. This gradual refinement, from less defined to more specialized representations, is consistent with how deep learning models learn hierarchically. Early layers capture basic signal properties, while deeper layers integrate these into more complex and meaningful patterns.

\begin{figure}[h] 
    \centering
    \includegraphics[width=0.9\textwidth]{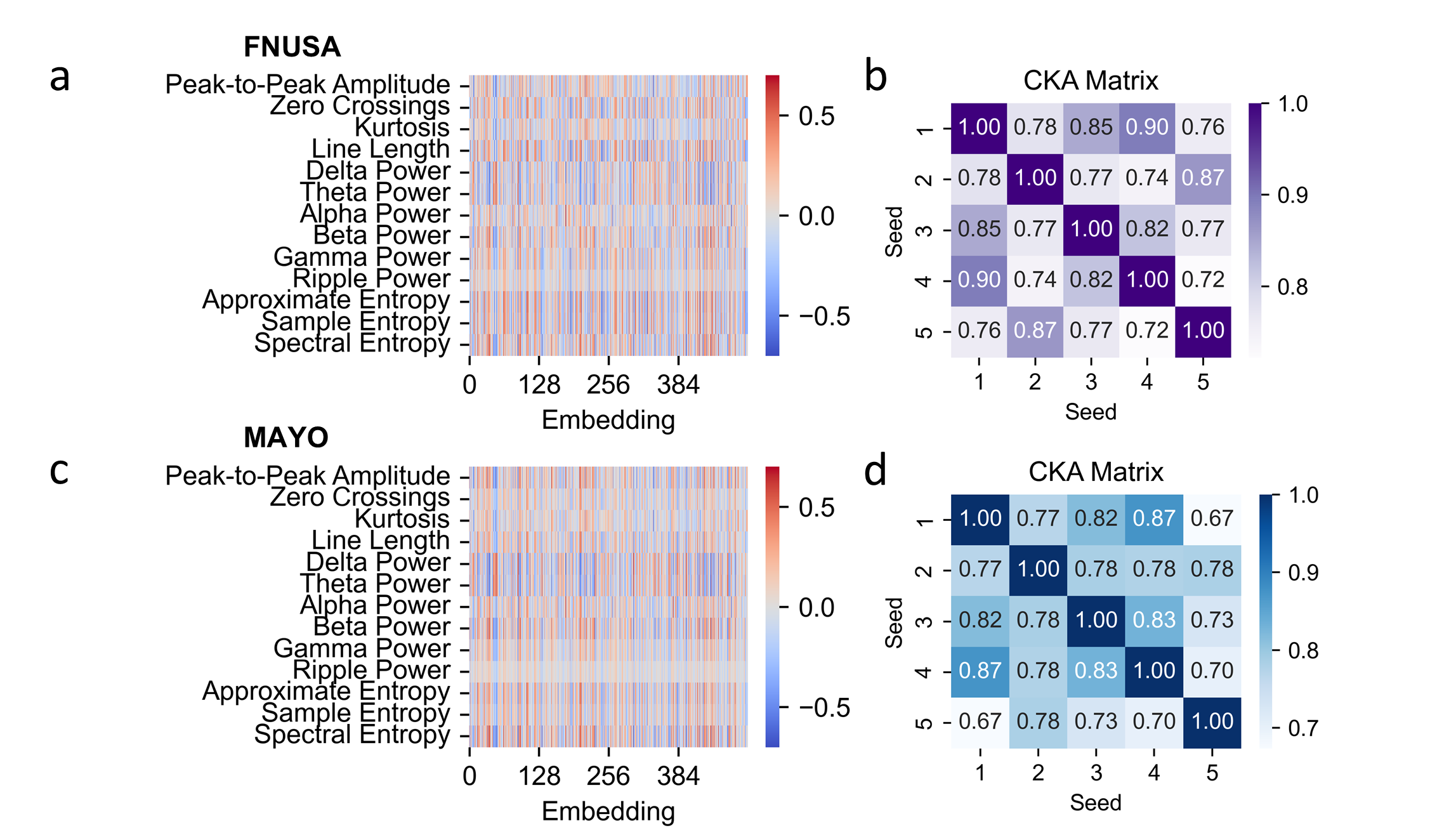} 
    \caption{\textbf{Interpretability analysis of EpiNT.} (a) Cosine similarity between EpiNT's extracted features and conventional EEG features on the FNUSA dataset. (b) CKA matrix illustrating feature consistency across different pre-training runs on the FNUSA dataset. (c) Cosine similarity between EpiNT's extracted features and conventional EEG features on the MAYO dataset. (d) CKA matrix illustrating feature consistency across different pre-training runs on the MAYO dataset. The CKA results demonstrated high cross-run alignment, with mean CKA scores of 0.798 ± 0.059 for FNUSA and 0.773 ± 0.060 for MAYO.}
    \label{fig:interpretablitiy}
\end{figure}

To enhance the interpretability of the learned deep learning features, we compared them to 13 commonly used EEG features spanning the spectral, time, and entropy categories. Our analysis, presented in Figure \ref{fig:interpretablitiy}a and \ref{fig:interpretablitiy}c, revealed that the model's deep representations share some similarities with these traditional features. For the FNUSA dataset, we found that 6.0\% of the time-domain features, 9.1\% of the spectral-domain features, and 11.9\% of the entropy features exhibited an absolute cosine similarity exceeding 0.5. Within the spectral-domain features for FNUSA, Delta power and Theta power accounted for 3.1\% and 2.9\% of the instances with an absolute cosine similarity greater than 0.5, respectively. For the MAYO dataset, 2.7\% of the time-domain features and 14.6\% of the spectral-domain features demonstrated an absolute cosine similarity greater than 0.5. For the entropy features of MAYO, the proportion of features with an absolute cosine similarity above 0.5 was only 5.4\%. This suggests that while EpiNT learns complex representations, these representations do encapsulate some of the information captured by established physiological measures.

Finally, we assessed the stability of the pre-trained model's learned features across different runs. We conducted five independent pre-training runs, each starting with different random initializations. We then quantified the similarity of the features extracted from the final layer of each run using Centered Kernel Alignment (CKA) as the metric. As shown in Figure \ref{fig:interpretablitiy}b and \ref{fig:interpretablitiy}d, the CKA scores between different runs were consistently high (around 0.7 to 0.9). These high scores indicate that despite the inherent randomness in deep learning training, the self-supervised pre-training process consistently converges to similar and stable feature representations. This robustness is crucial, as it means the model's ability to extract meaningful features is not overly dependent on the initial random setup.

\subsection{Evaluation of Designing Choices during Pre-training}
To optimize the pre-training strategy, we systematically examined the effects of vector quantization parameters, masking ratios, and key model components, aiming to inform future studies on pre-training models for neural signals.

We assessed how the configuration of vector quantization, used to map continuous signal segments into discrete prototypes, affects model performance. Specifically, we varied the codebook size (number of prototypes in total) and the number of quantizers (prototypes per segment). As shown in Supplementary Figures 2 and 3, performance gains from increasing the codebook size (from 256 to 2048) were inconsistent across tasks. Notably, with small codebooks (e.g., size 16), adding quantizers improved performance significantly, but returns diminished with larger codebooks. These findings suggest that optimal performance depends on balancing the representational capacity of the codebook with the complexity of the quantizer configuration.

Next, we analyzed how masking ratios during pre-training influence downstream task performance. Higher masking led to increased initial loss and slower convergence (Supplementary Figure 4), and generally impaired downstream results, particularly for the FNUSA and MAYO datasets (e.g., FNUSA F1 score dropped from 0.805 to 0.631; see Supplementary Figure 5). An exception was the HUH dataset, which showed marginal improvement. Overall, moderate masking supports feature learning, while excessive masking impairs signal reconstruction.

To evaluate the contributions of individual components in our EpiNT framework, we conducted ablation studies (Supplementary Figure 7). Beyond performance variations introduced by different positional encoding (which informs the model about the sequence order of signal patches) and masking strategies (how the model hides patches of the signal), our investigation revealed more insights from contrasting pre-training paradigms and the design of the vector quantizer.
We first compared our vector quantization-based pre-training (Model 1) against a direct reconstruction paradigm (Model 6). The direct reconstruction paradigm, which simply learns to rebuild the input signal using an L2 loss without vector quantization,  demonstrably underperformed our proposed method. Further, within the vector quantization framework, we found that mapping the data in the frequency domain (Model 1) consistently outperformed mapping in the time domain (Model 7). This highlights that the most informative features in neural signals, particularly those relevant to epilepsy and often found in higher frequencies, are better captured when processed in the frequency domain. Moreover, the inherently low signal-to-noise ratio of neural signals suggests that direct reconstruction may hinder the model's ability to learn robust and useful feature representations.

\subsection{Results for Comparisons}
We conducted a series of experiments to understand the influence of different fine-tuning methods and benchmark EpiNT. We assessed three fine-tuning strategies for EpiNT: linear-probing (training only the classification head), last-layer tuning (updating the final Transformer layer and head), and full-parameter tuning (updating all model weights). See Supplementary Note 1 for more training details. To assess the benefit of EpiNT, we compared it with both randomly initialized models and other Transformer-based pre-training models. Additionally, we conducted out-of-the-box evaluations to examine how different pre-training datasets influence transferability.

First, EpiNT consistently outperformed randomly initialized models, confirming that pre-training significantly enhances performance (Supplementary Tables 4, 5, 6, 7, and 8). Notably, in linear-probing, EpiNT's frozen backbone still produced rich features for classification, outperforming other pre-trained models in terms of F1 score (Table \ref{tab: compare, pre-trained}), though sometimes at the cost of lower specificity. This suggests that EpiNT captures generalizable features, even when the classifier is shallow. However, in larger and more diverse datasets (e.g., HUH), its performance was more limited under linear-probing, highlighting that complex data may require more flexible fine-tuning to fully leverage the pre-trained representations. We also observed that models employing reconstruction-based pre-training objectives (e.g., MAE-EEG, Brant, MOMENT, PatchTST) generally underperform those utilizing vector-quantized approaches, such as VQ-MTM and EpiNT, in linear-probing setting.

Second, the comparison of fine-tuning strategies highlighted important trade-offs between adaptability and flexibility. While EpiNT achieved the highest F1 scores in four out of six evaluation tasks, namely TASMC-UCLA (0.925$\pm$0.009), CHB-MIT (0.913$\pm$0.003), FNUSA (0.886$\pm$0.005), and MAYO (0.965$\pm$0.003), its performance declined under full-parameter fine-tuning relative to linear-probing. Linear-probing retains most of the pre-trained knowledge and is computationally efficient. Last-layer tuning offers greater flexibility while still preserving general representations. In contrast, full-parameter tuning enables more task-specific adaptation and generally yields stronger performance, but at the cost of increased risk of overfitting and computational demand. Signs of overfitting suggest that more extensive adaptation does not necessarily translate into better generalization (Supplementary Tables 9, 10, 11, and 12).

These results emphasize that fine-tuning depth should be matched to task complexity and data variability. For general clinical deployment with limited resources, efficient strategies like last-layer tuning may suffice; for specialized diagnostic tasks, full adaptation could be more beneficial, albeit with caution against overfitting.

\section{Discussions}
The experimental evaluation of the proposed EpiNT pre-training method yielded compelling results. Notably, EpiNT demonstrated favorable F1 scores, indicative of robust average performance, and good sensitivity under the linear-probing fine-tuning setting. This suggests the learned representations exhibit both generalizability and adaptability across a range of downstream tasks. While the specificity and NPV of EpiNT were comparatively modest in certain scenarios, these metrics generally fell within the performance range observed for other pre-trained models. Furthermore, these potential limitations can likely be mitigated through fine-tuning a larger number of parameters.

During the pre-training of EpiNT, while explicit annotations were not utilized to guide the learning process, we implicitly differentiated between signal sources (i.e., modality annotation) through modality-specific data pre-processing strategies. Specifically, scalp EEG was downsampled to 256 Hz and divided into 12-second segments, while iEEG was resampled to 1024 Hz and divided into 3-second segments. These pre-processing choices reflect how such data are typically handled in practice and help preserve the natural distribution of clinical recordings. As a result, the dataset includes inherent imbalances, such as a larger volume of scalp EEG data compared to iEEG and a predominance of inter-ictal data over ictal data. From a replicability perspective, data conforming to this natural distribution is readily accessible, facilitating the pre-training of larger-scale models with the acquisition of more data. Furthermore, from an application perspective, this modality-specific pre-processing may enable the model to better capture inherent modality-specific information, potentially leading to enhanced generalizability in real-world clinical settings. However, the implicit annotation of modality information could also limit the model's flexibility, potentially resulting in suboptimal performance when confronted same modality with variations in processing pipelines.

Despite our dataset comprising over 2700 hours of EEG and iEEG data from 1199 individuals, the significant inter-subject variability in brain activity and imbalanced data distribution, presents challenges for robust generalization across diverse populations. The observation that models trained from scratch outperformed pre-trained models on the CHB-MIT dataset suggests a potential domain shift between the pre-training and downstream datasets. This domain shift could be due to differences in recording equipment, clinical environments, or patient characteristics such as age, gender, or medical history. Moreover, we did not observe a strictly monotonic relationship between the volume of pre-training data and downstream task performance, unlike in \cite{jiang2024large}. This may be due to the nature of clinical EEG data, which often has a lower signal-to-noise ratio because of artifacts and reflects spontaneous rather than evoked brain activity (e.g., P300, motor imagery, emotion recognition). Meanwhile, while laboratory-based EEG experiments often yield balanced data distributions, the inherent imbalance characteristic of clinical data may hinder the performance of pre-trained models on certain tasks, such as seizure detection and prediction.

Comparisons with other pre-trained models reveal substantial variation in architecture and training objectives, with reconstruction-based approaches generally underperforming vector-quantized methods (Supplementary Note 2). Beyond the architectures included in our direct comparison, other notable models in the field offer valuable insights, particularly regarding the handling of input modality. Our study adopted a single-channel modeling approach, which simplifies the processing of diverse neural signals that often come with inconsistent montages. In contrast, models such as MMM and LaBraM are explicitly designed for multi-channel scalp EEG, enabling them to explicitly leverage inter-channel dependencies. Due to the heterogeneity of downstream tasks in our evaluation, a direct performance comparison with these multi-channel models was not conducted. However, their reported performance on scalp EEG-based tasks such as emotion recognition (achieving over 90\% accuracy) and anomaly EEG classification (achieving over 80\% accuracy) highlights the potential benefits of multi-channel modeling for specific applications. These observations offer valuable insights for the future design of pre-trained models for neural signals: (1) the development of more suitable embedding layers to ensure the effective abstraction of neural signal patches within Transformer architectures; (2) the enhanced utilization of the inherent spatial information within neural signals; and (3) the careful consideration of downstream task construction for better evaluating the capabilities of pre-trained models.

Our investigation into MAE and VQ paradigms for pre-training reveals several modality-specific insights that are critical for effective neural signal representation learning. We found that, in single-codebook VQ, there is an optimal codebook size. A codebook that is too small constrains representational capacity and increases quantization distortion, leading to the loss of important signal details. Conversely, excessively large codebooks introduce a higher risk of overfitting, especially when pre-training data is limited. A key finding is that using multiple smaller codebooks outperforms using a single large one. These codebooks appear to complement each other, with each capturing different aspects of the data, resulting in a more comprehensive and balanced representation. This approach is analogous to combining several simple models to enhance overall performance. As a form of information compression, the effectiveness of VQ is closely tied to codebook design, which should reflect the intrinsic information density of neural recordings. Furthermore, we observed that increasing the mask ratio in MAE degrades performance on neurophysiological data, in contrast to trends observed in image-based tasks \cite{he2022masked}. This difference likely stems from how information is distributed: images often contain spatially redundant patterns, allowing masked regions to be inferred from nearby context, whereas EEG and iEEG signals—particularly in epilepsy—often include brief, sparse events such as epileptic spikes or high-frequency bursts. These events are critical for diagnosis and are easily lost with high masking ratios. Overall, our findings emphasize that pre-training strategies must be carefully adapted to match the unique characteristics of neural signals, especially when they carry medically important information.

As a pre-trained deep learning model for EEG and iEEG analysis, EpiNT offers potential for clinical translation and future epilepsy research. A key immediate advantage is its robust generalization across different individuals. In clinical practice, patient-specific data are often limited. Unlike models trained from scratch on small datasets, EpiNT leverages its extensive pre-training on diverse data to provide reliable analysis for new patients, helping to address the challenge of inter-patient variability. This attribute is particularly valuable in data-scarce scenarios, such as the study of rare seizure types or specific epileptic syndromes, where the model's foundational knowledge can enable effective analysis even with limited task-specific data. Looking forward, the development of EpiNT can follow two complementary paths: specialization and generalization. For specialization, the base model can be fine-tuned for specific clinical applications and compressed into a lightweight version for real-time monitoring on resource-constrained devices. For generalization, integrating multi-modal data such as neuroimaging (e.g., fMRI) and genomics into pre-training of EpiNT may facilitate the discovery of novel biomarkers and provide a more comprehensive understanding of epilepsy. These efforts may strengthen the potential of EpiNT to inform both the scientific understanding and clinical management of epilepsy.

\section{Methods}
\subsection{Datasets}
We assembled a diverse collection of epileptic EEG datasets to pre-train a Transformer model, incorporating both scalp and intracranial recordings \cite{li2021neural} \cite{fedele2017resection} \cite{ds003555} \cite{ds003844} \cite{ds003876} \cite{ds004100} \cite{dimakopoulos2022information} \cite{zhang2022refining} \cite{veloso2017big} \cite{shah2018temple} \cite{detti2020eeg} \cite{nasreddine2021epileptic}. This dataset compilation was designed to improve generalizability and robustness across a variety of epilepsy-related tasks. Given the differences in acquisition montages among SEEG, ECoG, and scalp EEG recordings, we partitioned the multi-channel data into single-channel epochs. Each single-channel epoch is treated as an independent training instance, thereby mirroring real-world clinical scenarios where the availability of channels may vary.

To assess the performance of our pre-trained model on epilepsy-related tasks, we defined five classification tasks across six datasets. For datasets lacking conventional names, we adopted the abbreviation of the releasing institution for nomenclature. The TASMC-UCLA dataset, comprising overnight sleep iEEG recordings from 25 subjects with epilepsy \cite{falach2024annotated}, was utilized for SOZ localization by classifying signals into SOZ and non-SOZ activity. The CHB-MIT scalp EEG dataset includes data from 22 pediatric patients with intractable seizures \cite{shoeb2009application}. It was employed for seizure prediction, distinguishing between inter-ictal and pre-ictal signals. The CUK-IMHANS dataset contains scalp EEG from 11 pediatric epilepsy patients and 10 healthy controls; only the patient data were used for this study \cite{fasil2021scalp}. CUK-IMHANS was used for IED detection, classifying scalp EEG recordings based on the presence or absence of IEDs. The HUH dataset \cite{stevenson2019dataset}, consisting of scalp EEG data from 79 neonates from Helsinki University Hospital, was used for seizure detection, classifying ictal and inter-ictal activity. The FNUSA and MAYO datasets \cite{nejedly2020multicenter}, containing iEEG from 14 and 25 patients, respectively, were employed for the classification of pathological and physiological brain activity. For all downstream tasks, we adopted a rigorous cross-subject split protocol: 80\% of the patients were assigned to the training set for fine-tuning the pre-trained model, while the remaining 20\% were reserved for testing. In alignment with our pre-training approach, each downstream evaluation also used single-channel epochs as independent input instances. 

We adhered to the Transparent Reporting of a multivariable prediction model for Individual Prognosis Or Diagnosis for Artificial Intelligence (TRIPOD+AI) \cite{collins2024tripod+} reporting guideline during the preparation of this manuscript (Supplementary Note 3). 

\subsection{Problem Formulation}
We consider the problem of learning representations from multi-channel EEG/iEEG time series data. Let $\mathcal{D}={X_1, X_2, .., X_N}$ denote a dataset of $N$ EEG/iEEG recordings, where each $X_i \in \mathbb{R}^{{C_i}\times T}$ represents a multi-channel time series with $C_i$ channels and $T$ time points. Our objective is to learn a function that effectively captures the underlying dynamics and patterns within these recordings.

Clinically relevant epileptic features, such as HFOs and IEDs, are typically transient and localized, often appearing in only a subset of recording channels. This characteristic, coupled with the variability in acquisition montages across datasets, presents a significant hurdle for cross-dataset generalization. To address this challenge, we employ a channel-independent modeling approach. we decompose each multi-channel recording $X_i$ into a set of single-channel time series. Specifically, for each $X_i$, we extract $C_i$ single-channel signals, denoted as $\{x_i^1, x_i^2, ..., x_i^{C_i}\}$ where each $x_i^{c} \in \mathbb{R}^{{1}\times T}$ represents the time series from the $c$-th channel of the recording. This decomposition allows us to focus on learning representations at the individual channel level, mitigating the problem of montage inconsistency acrossing different modalities.

Formally, we seek to learn a mapping function $f_{\theta}: \mathbb{R}^{1\times T} \to \mathbb{R}^{1\times d}$, parameterized by $\theta$, which maps each single-channel time series $x_i^{c}$ to a $d$-dimensional representation $H_i^{c} = f_{\theta}(x_i^{c})$. The parameter $\theta$ is learned by minimizing a specific loss function $\mathcal{L}$ defined over the dataset of single-channel signals. Thus, the problem we address can be formulated as:
\begin{equation}
    \min_{\theta}\ \mathcal{L}(\theta) = \min_{\theta}\ \frac{1}{\sum_{i=1}^N C_i} \sum_{i=1}^N \sum_{c=1}^{C_i} \mathcal{L}(f_{\theta}(x_i^c)) 
\end{equation}

\subsection{Model Architecture}
\begin{figure}[h] 
    \centering
    \includegraphics[width=0.9\textwidth]{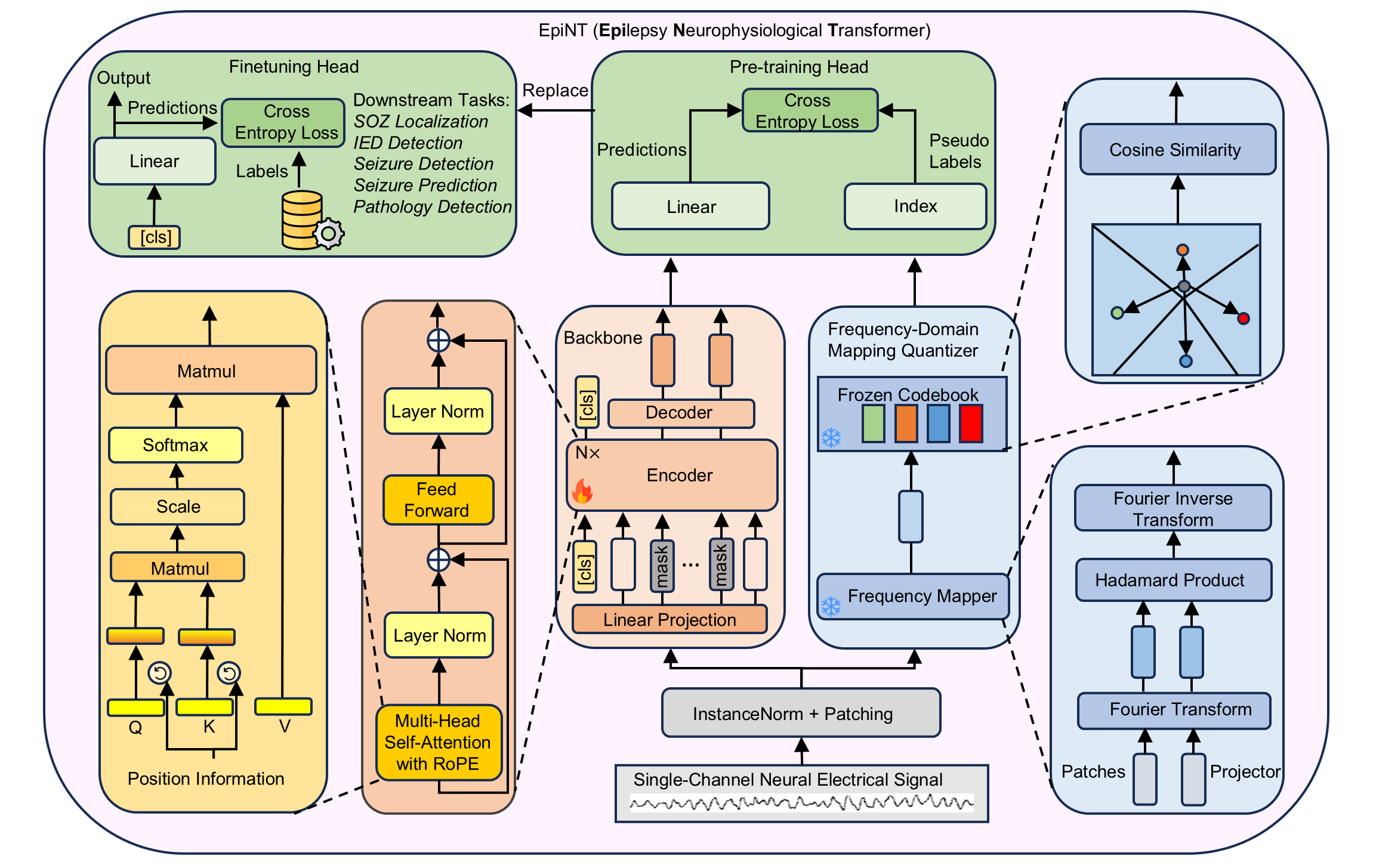} 
    \caption{\textbf{Overall architecture of EpiNT.} Single-channel electrophysiological signal is processed by instance normalization and patch embedding before being fed into the backbone. During the pre-training stage, the backbone consists of stacked Transformer layers acting as the encoder and a lightweight linear layer serving as the decoder. During fine-tuning, the learnable classification token ([cls] token) pre-appended to the input sequence and used for classification. The quantizer, initialized randomly at the beginning of training, remains frozen throughout both pre-training and fine-tuning. It employs a frequency mapper and cosine similarity to generate pseudo-labels.}
    \label{fig:method} 
\end{figure}

To effectively leverage the abundance of unlabeled neurophysiological data and address the inherent discrepancies across different electrophysiological modalities, we employ a modeling approach combining MAE and VQ. Figure \ref{fig:method} illustrated the model architecture of proposed EpiNT. MAE works by masking parts of the input signal, which encourages the model to learn the intrinsic information in brain activity. In addition, variations in amplitude and SNR exist across different modalities, such as EEG and iEEG, posing challenges for cross-modal modeling. VQ helps address this issue by mapping continuous signals into a simplified and discrete space, reducing these differences and making the model more stable and robust. 

The proposed EpiNT architecture consists of an initial embedding layer followed by a stack of Transformer layers serving as encoders. For pre-training, EpiNT employs a frequency-domain mapping quantizer and a lightweight linear layer as the decoder, utilizing a masking strategy. During fine-tuning, a linear layer is connected to the output of the encoder, and the masking strategy and frequency-domain mapping quantizer are disabled.

For the embedding layer, EpiNT takes a batch of single-channel neurophysiological signals as input, where each signal is a time series $x_i^{c} \in \mathbb{R}^{{1}\times T}$. The embedding layer (see Section \ref{sec: embedding layer} for more details) first divides each single-channel signal into non-overlapping patches of length $L = 256$ time points. Each patch is then linearly projected into a $d$-dimensional embedding space using a learnable embedding matrix. Before being fed into the Transformer encoder, a classification token $[\text{cls}]$ is prepended to the sequence of patch embeddings. The resulting sequence, consisting of the $[\text{cls}]$ token followed by the patch embeddings, serves as the input to the Transformer encoder.

The encoder backbone, denoted as $f_{\theta}$, comprises a stack of Transformer layers that takes embedded patches as input and outputs a sequence of latent representations (see Section \ref{sec: Encoder} for more details). Since EEG and iEEG are inherently sequential, we utilize Rotary Positional Embedding (RoPE) \cite{su2024roformer} to integrate positional information directly within the self-attention mechanism. This approach is chosen due to RoPE's capability to effectively capture the relative positional relationships inherent in the temporal dynamics of EEG and iEEG signals through its rotary operations (see Section \ref{sec: MHSA-RoPE} for more details). The output of the last Transformer layer is the sequence of latent representations $H_i^{c}$.

During the pre-training stage, to encourage the learning of robust representations, a masking strategy is applied to the embedded patches. Specifically, a proportion $m$ of the patches, where $0 < m < 1$, is randomly masked and subsequently replaced with noise sampled from a normal distribution with a mean of $0$ and a standard deviation of $0.1$, mimicking the inherent noise characteristics of neural signals. While the entire sequence of embedded patches $Z_i^c$ is fed into the encoder, the loss computation is performed exclusively on the masked patches. Following the encoder's processing, a lightweight decoder $g_{\phi}$ is employed to reconstruct the masked embedded patches. This decoder takes the latent representations $H_i^c$ from the encoder as input and consists of a single linear layer that projects these representations back to the original patch dimension, as formulated by:
\begin{equation}
    \hat{Z}_i^c = g_{\phi}(H_i^c) = H_i^c W_O
\end{equation}
where $W_{O} \in \mathbb{R}^{{d}\times L}$ represents a learnable weight matrix. 

To facilitate self-supervised pre-training, VQ is utilized to generate pseudo-labels. In this paper, we proposed a novel frequency domain mapping quantizer, as detailed in Section \ref{sec: Frequency Domain Mapping Quantizer}. We employ $Q$ distinct quantizers, indexed by $q \in \{1, 2, ..., Q\}$. Each quantizer $q$ is associated with a fixed, non-trainable codebook $C^{q} = \{c_1^q, c_2^q, ..., c_{K_q}^q\}$, where $K_q$ denotes the codebook size of the $q$-th quantizer. Each codebook vector $c_{k}^{q} \in \mathbb{R}^{1\times d}$ resides in a $d$-dimensional embedding space. Given an input sequence partitioned into $n$ patches, the reconstructed representation of the $j$-th patch by the decoder is denoted as $\hat{z}_j \in \hat{Z}^{c}_{i}$. For each quantizer $q$, a corresponding linear layer $W_q \in \mathbb{R}^{d \times K_q}$ is applied to the reconstructed representation $\hat{z}_j$ to predict the pseudo-label $l_j^q \in \{1, 2, ..., K_q\}$. With $Q$ quantizers, the overall pre-training loss is computed as the average of the individual quantizer losses. Specifically, the reconstruction loss is defined as:
\begin{equation}
    \mathcal{L}_{rec}(P_i^c, \hat{Z}_i^c) = -\frac{1}{Qn} \sum_{q=1}^{Q} \sum_{j=1}^{n} \log(\text{softmax}(\hat{z}_j W_q)_{l_j^q})
\end{equation}
where $\text{softmax}(\hat{z}_j W_q)_{l_j^q}$ denotes the probability of the $l_j^q$-th class predicted by the classifier $W_q$ for the reconstructed patch $\hat{z}_j$. This multi-quantizer approach is designed to encourage the encoder to learn more robust and discriminative representations.

In the fine-tuning stage, the pre-trained EpiNT encoder is adapted for downstream classification tasks. A linear classifier, denoted as $W_{[\text{cls}]}$, is applied to the encoder's output corresponding to the classification token, $f_\theta(z_{[\text{cls}]})$, to predict the class label at the sequence level:
\begin{equation}
    y = \text{softmax}(f_\theta{(z_{[\text{cls}]})} W_{[\text{cls}]})
\end{equation}
The model is then fine-tuned using the cross-entropy loss function. The fine-tuning head typically comprises a single linear layer with an output dimensionality matching the number of classes in the target task. The input dimension of this layer corresponds to the feature vector dimension of the classification token output from the pre-trained encoder. Alternatively, the pre-trained EpiNT backbone can be frozen and utilized as a feature extractor. In this scenario, the feature vector from the classification token can be fed into a separate classifier, such as a Support Vector Machine (SVM), for downstream classification.

\subsection{Embedding Layer for EpiNT} \label{sec: embedding layer}
Specifically, for each single-channel time series $x_i^{c} \in \mathbb{R}^{{1}\times T}$, we use $\text{InstanceNorm}$ to normalize it independently, effectively eliminating individual differences. Then, we divide it into non-overlapping patches of length $L = 256$ time points. This results in a sequence of patches, denoted as $P_i^c = \{p_1, p_2, ..., p_n\}$, where each patch $p_j \in \mathbb{R}^{{1}\times L} $, $j \in \{1, 2, ..., n\}$ and $n = \lfloor T/L \rfloor$ is the number of patches. Each patch $p_j \in \mathbb{R}^L$ is then linearly projected into a $d$-dimensional embedding space using a learnable embedding matrix $E \in \mathbb{R}^{L \times d}$:
\begin{equation}
  z_j = p_j E  
\end{equation}
Before feeding the patch embeddings into the Transformer encoder, a special classification token $[\text{cls}]$ is prepended to the sequence. This $[\text{cls}]$ token is a learnable vector $z_{[\text{cls}]} \in \mathbb{R}^{1 \times d}$ that aggregates information from the entire sequence and is used for downstream classification tasks during fine-tuning. The input sequence to the Transformer encoder is therefore:
\begin{equation}
    Z^{c}_{i}=[z_{[\text{cls}]}, z_1, z_2, ..., z_N]    
\end{equation}
where $z_j$ represents the embedding of the $j$-th patch. These patch embeddings, along with the $[\text{cls}]$ token, form the input sequence to the Transformer encoder.

\subsection{Encoder for EpiNT} \label{sec: Encoder}
The encoder $f_{\theta}$ for backbone is a stack of Transformer layers, which takes the  masked and unmasked embedded patches as input and outputs a sequence of latent representations. Each Transformer layer processes the input sequence as follows. First, a multi-head self-attention mechanism with rotary positional encoding (MHSA-RoPE) is applied to capture the relationships between different embedded patches, which will illustrated in Section \ref{sec: MHSA-RoPE}. Then, a residual connection is added between the input and the output of the attention mechanism, followed by layer normalization:
\begin{equation}
    \text{AttnNorm}({Z}_i^c) = \text{LayerNorm}({Z}_i^c + \text{MHSA-RoPE}({Z}_i^c))
\end{equation}
Next, a feed-forward network (FFN) is applied to each embedded patch. The FFN consists of two linear layers with a ReLU activation function in between:
\begin{equation}
    \text{FFN}(x) = \text{ReLU}(x W_1) W_2
\end{equation}
where $W_1 \in \mathbb{R}^{{d}\times 4d}$ and $W_2 \in \mathbb{R}^{{4d}\times d}$ are learnable weight matrices. For simplicity, bias terms are omitted. Finally, another residual connection and layer normalization are applied:
\begin{equation}
    \text{Output} = \text{LayerNorm}(\text{AttnNorm}({Z}_i^c) + \text{FFN}(\text{AttnNorm}({Z}_i^c)))
\end{equation}
The output of the last Transformer layer is the sequence of latent representations $H_i^{c}$.

\subsection{Multi-Head Self-Attention with RoPE} \label{sec: MHSA-RoPE}
For pre-trained models designed for EEG and iEEG, positional information for sequence is typically incorporated by adding positional encodings to the input embeddings. However, in EpiNT, we employ RoPE to integrate positional information directly within the self-attention mechanism, since it can capture relative position information in the signals via rotary operation.

RoPE encodes positional information through rotation matrices. Given two transposed embedding vectors $z_m$, $z_n \in \mathbb{R}^{d}$ representing the $m$-th and $n$-th embedded patches respectively, the RoPE-augmented attention score between them is computed as follows:
\begin{equation}
        q_m^T k_n  = 
         (R_{\Theta, m}^d W_q z_m)^T (R_{\Theta, n}^d W_k z_n) = W_q^T z_m^T R_{\Theta, n-m}^d z_n W_k
\end{equation}
where $W_q$, $W_k \in \mathbb{R}^{d \times d}$ are the query and key transformation matrices, respectively, and $R^{d}_{\Theta, m}$ is the rotation matrix for position $m$ in a $d$-dimensional space. The rotation matrix $R^{d}_{\Theta m}$ is defined as a block diagonal matrix:
\begin{equation}
        R_{\Theta, m}^d = 
        \begin{bmatrix}
            \cos m\theta_1 & -\sin m\theta_1 & & \cdots & 0 \\
            \sin m\theta_1 & \cos m\theta_1 & \ddots & & \vdots \\
            \vdots &  & \ddots & \cos m\theta_{d/2} & -\sin m\theta_{d/2} \\
            0 & \cdots &  & \sin m\theta_{d/2} & \cos m\theta_{d/2}
        \end{bmatrix}
    \label{eq:rotation matrix}
\end{equation}
where $\Theta = \{\theta_i = 10000^{-2(i-1)/d}, i \in [1, 2, ..., d/2] \}$ are pre-defined rotation angles. As shown in equation (\ref{eq:rotation matrix}), the attention score depends only on the relative positional difference $n-m$, which allows the model to effectively capture relative temporal information in EEG signals. 

To further enhance the model's capacity, we employ multi-head self-attention. Instead of performing a single attention operation, we project the query, key, and value vectors into $h$ different subspaces using learnable projection matrices $W_q^u$, $W_k^u$, $W_v^u \in \mathbb{R}^{d \times d_k}$, where $u\in \{1, 2, ..., h\}$ and $d_k = d/h$. For the first layer of the encoder ($l=1$), the input is $Z^{0} = Z^{c}_{i}$. For subsequent layers ($l>1$), the input is the output of the previous layer, denoted as $Z^{l-1}$. The RoPE-augmented attention is then computed independently for each head:
\begin{equation}
    \text{Attn}_u(Q^{l}, K^{l}, V^{l}) = 
     \text{Softmax}(\frac{(R_{\Theta, :}^d Q^{l})^T (R_{\Theta, :}^d K^{l})}{\sqrt{d_k}}) V^{l}
\end{equation}
where $Q^{l}=W_q^uZ^{l-1}$, $K^{l}=W_k^uZ^{l-1}$, and $V^{l}=W_v^uZ^{l-1}$ are the query, key, and value matrices for head $u$, respectively, and $R_{\Theta, :}^d$ represents the application of RoPE to all input positions. The outpus of all heads are then concatenated and linearly projected back to the original dimension:
\begin{equation}
        \text{MHSA-RoPE}(Q^{l}, K^{l}, V^{l})    = 
                                                \text{Concat}(\text{Attn}_1, \text{Attn}_2, ..., \text{Attn}_h) W_O
\end{equation}
where $W_O \in \mathbb{R}^{(h \cdot d_k) \times d} $ is the projection matrix. By using MHSA-RoPE, EpiNT can effectively capture both local and global temporal dependencies in the EEG/iEEG signals while efficiently incorporating positional information of patches.

\subsection{Frequency Domain Mapping Quantizer} \label{sec: Frequency Domain Mapping Quantizer}
Scalp EEG and iEEG exhibit inherent differences in frequency content due to variations in recording techniques. Scalp EEG signals suffer significant attenuation of high-frequency components, while iEEG captures richer high-frequency information. Consequently, a key pre-training challenge is accurately characterizing the frequency characteristics of different electrophysiological signals during quantization. Moreover, the transient and localized nature of epilepsy-related features further necessitates a focus on local temporal information during this process. To address these issues, we propose a novel frequency domain mapping quantizer, which is randomly-initialized and non-trainable. Unlike traditional random projection VQ methods that project signals via matrix multiplication, our quantizer operates circular convolution with input signal patches in the time domain.
Circular convolution in the time domain is equivalent to the Hadamard product of the Fourier transforms of the corresponding signals in the frequency domain.
This implies that our quantizer explicitly processes frequency domain information and implicitly captures local temporal features without requiring additional frequency feature engineering. 

Specifically, given an original patch $p_{j} \in \mathbb{R}^{L}$, we first perform a frequency domain transformation using a randomly initialized vector $h_{proj} \in \mathbb{R}^{L}$. This transformation involves computing the Discrete Fourier Transform (DFT) of both the patch $p_{j}$ and the random vector $h_{proj}$, followed by a Hadamard product in the frequency domain:
\begin{equation}
        \tilde{p}_j = \mathcal{F}^{-1}(\mathcal{F}(p_j) \odot \mathcal{F}(h_{proj}))
    \end{equation}
where $\mathcal{F}$ and $\mathcal{F}^{-1}$ denote the DFT and Inverse Discrete Fourier Transform (IDFT), respectively, and $\odot$ represents the Hadamard product (element-wise multiplication). This operation is equivalent to circular convolution in the time domain.
To measure the distance between the transformed patch and the codebook vectors, we employ cosine similarity. To establish a direct correspondence between cosine similarity and Euclidean distance, both the transformed patch $\tilde{p}_{j}$ and each codebook vector $c_k \in \mathbb{R}^{L}$ are normalized to unit length:
\begin{equation}
    \hat{p}_j = \frac{\tilde{p}_j}{||\tilde{p}_j||_2}, \quad \hat{c}_k = \frac{c_k}{||c_k||_2}
    \label{eq:norm}
\end{equation}
The cosine similarity between the normalized transformed patch $\tilde{p}_{j}$ and each normalized codebook vector $\tilde{c}_{k}$ is then computed as:
\begin{equation}
    \text{cos}(\hat{p}_j, \hat{c}_k) = \hat{p}_j \cdot \hat{c}_k
    \label{eq:cos_sim}
\end{equation}
When both vectors are normalized to unit length, maximizing cosine similarity is equivalent to minimizing the Euclidean distance. Finally, the index of the codebook vector yielding the highest cosine similarity (or equivalently, the smallest Euclidean distance) is assigned as the pseudo-label:
\begin{equation}
    l_j = \arg\max_k \text{cos}(\hat{p}_j, \hat{c}_k) = \arg\min_k ||\hat{p}_j - \hat{c}_k||_2
    \label{eq:quant}
\end{equation}
This approach not only better accommodates the disparate frequency characteristics of different EEG modalities but also aligns with the localized nature of epileptic features.

\subsection{Data Preprocessing}
A consistent preprocessing pipeline was applied to all data used for both pre-training and downstream tasks. Specifically, all scalp EEG recordings were resampled to 256 Hz, while all iEEG recordings (including ECoG and SEEG) were resampled to 1024 Hz. This modality-specific resampling strategy is determined by the distinct frequency characteristics of scalp and iEEG signals. 256 Hz is typically sufficient for capturing the relevant frequency content in scalp EEG, while 1024 Hz is necessary to preserve the high-frequency components present in iEEG. To maintain consistent input dimensions for the models, the data were segmented into epochs of 12 seconds for scalp EEG and 3 seconds for iEEG, resulting in 3072 time points per epoch for both modalities.

Furthermore, bad channels were identified and removed based on channel information provided within each dataset. To mitigate power line interference, a 50/60 Hz notch filter was applied to all recordings. Beyond these steps, no further pre-processing steps were implemented to avoid introducing potential biases or distorting the original signal characteristics.

\subsection{Exploration of Model Architecture and Pre-training Configurations}
To systematically investigate the impact of various pre-training factors on downstream task performance, including pre-training dataset modality, codebook size, number of quantizers, masking ratio, and dataset size, we conducted a series of targeted experiments. To quantify the respective contributions of different EEG modalities during pre-training, we trained two additional baseline models: a scalp-only model, pre-trained exclusively on scalp EEG data, and an intracranial-only model, pre-trained solely on iEEG data. To determine the influence of the quantizer codebook size on the model's generalization capabilities, we evaluated EpiNT with four distinct codebook sizes: 256, 512, 1024, and 2048. We further explored the effect of employing multiple quantizers on generalization performance by conducting experiments with varying numbers of quantizers under two codebook size configurations: a larger codebook with a size of 512 and a smaller codebook with a size of 16. To assess the impact of the masking ratio on model performance, we performed experiments using mask ratios ranging from 0.3 to 0.7. We also investigated the effect of pre-training dataset size on model performance by pre-training EpiNT on varying proportions of the total pre-training data: 40\%, 60\%, 80\%, and 100\%.

To gain deeper insights into the contributions of individual components within our EpiNT model, we conducted ablation studies. These studies involved systematically modifying or removing specific elements of the model architecture to assess their individual impact on the overall performance. Specifically, we examined the following:
\begin{itemize}
    \item Positional encoding method: We compared learnable positional embeddings (where position information is learned directly from data), sinusoidal positional encodings (fixed, non-trainable functions based on sine and cosine waves), and RoPE.
    \item Masking token strategy: We evaluated learnable token masking (where masked tokens are represented by a learnable embedding), all-zero token masking (where masked tokens are represented by zero vectors), and random noise masking.
    \item Pre-training paradigm: We compared direct reconstruction using L2-norm(minimizing the L2-norm between the reconstructed and original input patches) against our employed quantization-based pre-training method.
    \item VQ projection method: We compared traditional random projection in the time domain (implemented via matrix multiplication) against our proposed frequency domain mapping quantizer.
\end{itemize}

To thoroughly investigate the impact of EpiNT's model architecture and pre-training configurations on generalization performance, we fine-tuned the model for each downstream task using a linear-probing approach. This means the pre-trained EpiNT model acted as a fixed feature extractor, with only the parameters of the linear classification head being optimized during fine-tuning.

\subsection{Interpretability Analysis for EpiNT}
To enhance the interpretability of the learned deep learning features within our proposed EpiNT model, we firstly employed visualization techniques. Specifically, we visualized representative electrophysiological signals, such as HFOs and spikes, alongside their corresponding feature representations extracted from different layers of EpiNT. Subsequently, we conducted a systematic comparison between EpiNT’s extracted deep features and a comprehensive set of 13 established EEG features, categorized into three domains: (1) Frequency-domain, (2) Time-domain, and (3) Entropy. The Time-domain features considered were: (i) Peak-to-Peak Amplitude, (ii) Zero Crossings, (iii) Kurtosis, and (iv) Line Length. The Frequency-domain features included: (i) Delta Power, (ii) Theta Power, (iii) Alpha Power, (iv) Beta Power, (v) Gamma Power, and (vi) Ripple Power. The Entropy features included: (i) Approximate Entropy, (ii) Sample Entropy, and (iii) Spectral Entropy. To quantitatively assess the similarity between EpiNT's learned representations and these established features, we computed the cosine similarity between the embedding of the $[\text{cls}]$ token from EpiNT’s final layer (position 0), which serves as a sequence-level representation of the input instance, and each feature. 

Furthermore, to evaluate the stability of EpiNT as a feature extractor, we performed five independent pre-training runs with distinct random seeds (0, 1, 2, 3, and 4). We then measured the feature consistency between the final layer representations of each run pair using CKA \cite{kornblith2019similarity}. Due to the computational cost associated with CKA, we randomly sampled 10\% of the FNUSA and MAYO dataset for this feature consistency analysis.

\subsection{Comparative Evaluation of Fine-Tuning Strategies and Model Performance} \label{Performance Benchmarking}
To evaluate the adaptability of different fine-tuning strategies under varying computational constraints, we investigated two additional approaches beyond linear-probing: last-layer tuning and full-parameter fine-tuning. In last-layer tuning, only the final Transformer layer and the classification head were updated during training, while the remaining layers were frozen. In full-parameter tuning, all model parameters were allowed to be updated. To benchmark the performance of our proposed EpiNT model, we compared it with five representative pre-trained models designed for neurophysiological signals and general time series analysis: MAE-EEG \cite{cai2024mae}, Brant \cite{zhang2024brant}, MOMENT \cite{goswami2024moment}, PatchTST \cite{nie2022time}, and VQ-MTM \cite{gui2024vector}. All these models were pre-trained on the same dataset and evaluated using the same downstream tasks as EpiNT. We further conducted an out-of-the-box comparison by pre-training the baseline models on the publicly available datasets originally used in their respective publications—specifically, BCI Competition IV Datasets 2a and 2b, and the PhysioNet EEG Dataset \cite{tangermann2012review, goldberger2000physiobank}. This enabled an assessment of each model’s inherent performance, albeit without access to their original pre-trained weights. Finally, to isolate the effect of pre-training, we compared EpiNT against randomly initialized models trained from scratch under the same fine-tuning settings. The randomly initialized baselines included ConFormer \cite{song2022eeg}, CSPNet \cite{jiang2024csp}, EEGInception \cite{santamaria2020eeg}, EEGNet \cite{lawhern2018eegnet}, TinySleepNet \cite{supratak2020tinysleepnet}, and ATCNet \cite{altaheri2022physics}.

Specifically, MAE-EEG, Brant, and VQ-MTM have demonstrated strong performance in pre-training on neuronphysiological data. In contrast, MOMENT and PatchTST represent well-established pre-training paradigms for broader time series domains, e.g. financial time series. 

To further quantify the effectiveness of the pre-training paradigm, we compared the performance of EpiNT with that of randomly initialized models trained from scratch under the same fine-tuning settings. This comparison isolates the contribution of pre-training itself and demonstrates its advantage in extracting transferable representations from large-scale neural signal data. Specifically, we selected ConFormer \cite{song2022eeg}, CSPNet \cite{jiang2024csp}, EEGInception \cite{santamaria2020eeg}, EEGNet \cite{lawhern2018eegnet}, TinySleepNet \cite{supratak2020tinysleepnet}, and ATCNet \cite{altaheri2022physics} as representative baselines for comparison. Among these models, ATCNet utilizes a Transformer architecture, while the others are based on convolutional neural networks. Pairwise performance comparisons were conducted between EpiNT and each individual baseline model across all six downstream datasets.

\section*{Data Availability}
Download links for the datasets used to pre-train and fine-tune EpiNT are provided in the Supplementary Tables 2 and 3. 

\section*{Code Availability}
The pre-trained model weights and codes for processing pre-training data, training and evaluating EpiNT are available at: https://github.com/RunKZhang/EpiNT.

\bibliographystyle{unsrt}  
\bibliography{template}  

\begin{thebibliography}{10}

\bibitem{Rana2018inflammation}
Amna Rana and Alberto~E. Musto.
\newblock The role of inflammation in the development of epilepsy.
\newblock {\em Journal of Neuroinflammation}, 15(144), 2018.

\bibitem{thijs2019epilepsy}
Roland~D Thijs, Rainer Surges, Terence~J O'Brien, and Josemir~W Sander.
\newblock Epilepsy in adults.
\newblock {\em The Lancet}, 393(10172):689--701, 2019.

\bibitem{beghi2020epidemiology}
Ettore Beghi.
\newblock The epidemiology of epilepsy.
\newblock {\em Neuroepidemiology}, 54(2):185--191, 2020.

\bibitem{maturana2020critical}
Matias~I Maturana, Christian Meisel, Katrina Dell, Philippa~J Karoly, Wendyl D’Souza, David~B Grayden, Anthony~N Burkitt, Premysl Jiruska, Jan Kudlacek, Jaroslav Hlinka, et~al.
\newblock Critical slowing down as a biomarker for seizure susceptibility.
\newblock {\em Nature Communications}, 11(1):2172, 2020.

\bibitem{abou2022noninvasive}
Maurice Abou~Jaoude, Claire~S Jacobs, Rani~A Sarkis, Jin Jing, Kyle~R Pellerin, Andrew~J Cole, Sydney~S Cash, M~Brandon Westover, and Alice~D Lam.
\newblock Noninvasive detection of hippocampal epileptiform activity on scalp electroencephalogram.
\newblock {\em JAMA Neurology}, 79(6):614--622, 2022.

\bibitem{gunnarsdottir2022source}
Kristin~M Gunnarsdottir, Adam Li, Rachel~J Smith, Joon-Yi Kang, Anna Korzeniewska, Nathan~E Crone, Adam~G Rouse, Jennifer~J Cheng, Michael~J Kinsman, Patrick Landazuri, et~al.
\newblock Source-sink connectivity: A novel interictal {EEG} marker for seizure localization.
\newblock {\em Brain}, 145(11):3901--3915, 2022.

\bibitem{roy2019deep}
Yannick Roy, Hubert Banville, Isabela Albuquerque, Alexandre Gramfort, Tiago~H Falk, and Jocelyn Faubert.
\newblock Deep learning-based electroencephalography analysis: A systematic review.
\newblock {\em Journal of Neural Engineering}, 16(5):051001, 2019.

\bibitem{shi2023b2}
Shuiling Shi and Wenqi Liu.
\newblock {B2-ViT Net}: Broad vision transformer network with broad attention for seizure prediction.
\newblock {\em IEEE Transactions on Neural Systems and Rehabilitation Engineering}, 2023.

\bibitem{li2020epileptic}
Yang Li, Yu~Liu, Wei-Gang Cui, Yu-Zhu Guo, Hui Huang, and Zhong-Yi Hu.
\newblock Epileptic seizure detection in {EEG} signals using a unified temporal-spectral squeeze-and-excitation network.
\newblock {\em IEEE Transactions on Neural Systems and Rehabilitation Engineering}, 28(4):782--794, 2020.

\bibitem{lin2024vepinet}
Nan Lin, Weifang Gao, Lian Li, Junhui Chen, Zi~Liang, Gonglin Yuan, Heyang Sun, Qing Liu, Jianhua Chen, Liri Jin, et~al.
\newblock {vEpiNet}: A multimodal interictal epileptiform discharge detection method based on video and electroencephalogram data.
\newblock {\em Neural Networks}, 175:106319, 2024.

\bibitem{zhang2024pyhfo}
Yipeng Zhang, Lawrence Liu, Yuanyi Ding, Xin Chen, Tonmoy Monsoor, Atsuro Daida, Shingo Oana, Shaun Hussain, Raman Sankar, Aria Fallah, et~al.
\newblock {PyHFO}: Lightweight deep learning-powered end-to-end high-frequency oscillations analysis application.
\newblock {\em Journal of Neural Engineering}, 21(3):036023, 2024.

\bibitem{zhang2024oxcarnet}
Runkai Zhang, Rong Rong, Yun Xu, Haixian Wang, and Xiaoyun Wang.
\newblock {OxcarNet}: Sinc convolutional network with temporal and channel attention for prediction of oxcarbazepine monotherapy responses in patients with newly diagnosed epilepsy.
\newblock {\em Journal of Neural Engineering}, 21(5):056019, 2024.

\bibitem{he2016deep}
Kaiming He, Xiangyu Zhang, Shaoqing Ren, and Jian Sun.
\newblock Deep residual learning for image recognition.
\newblock In {\em Proceedings of the IEEE conference on computer vision and pattern recognition}, pages 770--778, 2016.

\bibitem{devlin2018bert}
Jacob Devlin.
\newblock Bert: Pre-training of deep bidirectional transformers for language understanding.
\newblock {\em arXiv preprint arXiv:1810.04805}, 2018.

\bibitem{banville2019self}
Hubert Banville, Isabela Albuquerque, Aapo Hyv{\"a}rinen, Graeme Moffat, Denis-Alexander Engemann, and Alexandre Gramfort.
\newblock Self-supervised representation learning from electroencephalography signals.
\newblock In {\em 2019 IEEE 29th International Workshop on Machine Learning for Signal Processing (MLSP)}, pages 1--6. IEEE, 2019.

\bibitem{cai2024mae}
Miao Cai and Yu~Zeng.
\newblock {MAE-EEG-Transformer}: A transformer-based approach combining masked autoencoder and cross-individual data augmentation pre-training for {EEG} classification.
\newblock {\em Biomedical Signal Processing and Control}, 94:106131, 2024.

\bibitem{wang2023brainbert}
Christopher Wang, Vighnesh Subramaniam, Adam~Uri Yaari, Gabriel Kreiman, Boris Katz, Ignacio Cases, and Andrei Barbu.
\newblock {BrainBERT}: Self-supervised representation learning for intracranial recordings.
\newblock {\em arXiv preprint arXiv:2302.14367}, 2023.

\bibitem{jiang2024large}
Wei-Bang Jiang, Li-Ming Zhao, and Bao-Liang Lu.
\newblock Large brain model for learning generic representations with tremendous {EEG} data in {BCI}.
\newblock {\em arXiv preprint arXiv:2405.18765}, 2024.

\bibitem{gui2024vector}
Haokun Gui, Xiucheng Li, and Xinyang Chen.
\newblock Vector quantization pretraining for {EEG} time series with random projection and phase alignment.
\newblock In {\em International Conference on Machine Learning}, pages 16731--16750. PMLR, 2024.

\bibitem{yi2024learning}
Ke~Yi, Yansen Wang, Kan Ren, and Dongsheng Li.
\newblock Learning topology-agnostic {EEG} representations with geometry-aware modeling.
\newblock {\em Advances in Neural Information Processing Systems}, 36, 2024.

\bibitem{zhang2024brant}
Daoze Zhang, Zhizhang Yuan, Yang Yang, Junru Chen, Jingjing Wang, and Yafeng Li.
\newblock Brant: Foundation model for intracranial neural signal.
\newblock {\em Advances in Neural Information Processing Systems}, 36, 2024.

\bibitem{kim2024towards}
Sung-Jin Kim, Dae-Hyeok Lee, Heon-Gyu Kwak, and Seong-Whan Lee.
\newblock Towards domain-free transformer for generalized {EEG} pre-training.
\newblock {\em IEEE Transactions on Neural Systems and Rehabilitation Engineering}, 2024.

\bibitem{hatlestad2023reliable}
Christoffer Hatlestad-Hall, Ricardo Bru{\~n}a, Mia Liljestr{\"o}m, Hanna Renvall, Kjell Heuser, Erik Taub{\o}ll, Fernando Maest{\'u}, and Ira~H Haraldsen.
\newblock Reliable evaluation of functional connectivity and graph theory measures in source-level {EEG}: How many electrodes are enough?
\newblock {\em Clinical Neurophysiology}, 150:1--16, 2023.

\bibitem{mitsuhashi2020effects}
Takumi Mitsuhashi, Masaki Sonoda, Hirotaka Iwaki, Aimee~F Luat, Sandeep Sood, and Eishi Asano.
\newblock Effects of depth electrode montage and single-pulse electrical stimulation sites on neuronal responses and effective connectivity.
\newblock {\em Clinical Neurophysiology}, 131(12):2781--2792, 2020.

\bibitem{saha2021progress}
Simanto Saha, Khondaker~A Mamun, Khawza Ahmed, Raqibul Mostafa, Ganesh~R Naik, Sam Darvishi, Ahsan~H Khandoker, and Mathias Baumert.
\newblock Progress in brain computer interface: Challenges and opportunities.
\newblock {\em Frontiers in Systems Neuroscience}, 15:578875, 2021.

\bibitem{mercier2022advances}
Manuel~R Mercier, Anne-Sophie Dubarry, Fran{\c{c}}ois Tadel, Pietro Avanzini, Nikolai Axmacher, Dillan Cellier, Maria Del~Vecchio, Liberty~S Hamilton, Dora Hermes, Michael~J Kahana, et~al.
\newblock Advances in human intracranial electroencephalography research, guidelines and good practices.
\newblock {\em Neuroimage}, 260:119438, 2022.

\bibitem{shakeshaft2022heterogeneity}
Amy Shakeshaft, Petroula Laiou, Eugenio Abela, Ioannis Stavropoulos, Mark~P Richardson, and Deb~K Pal.
\newblock Heterogeneity of resting-state {EEG} features in juvenile myoclonic epilepsy and controls.
\newblock {\em Brain Communications}, 4(4):fcac180, 2022.

\bibitem{sakakura2023developmental}
Kazuki Sakakura, Naoto Kuroda, Masaki Sonoda, Takumi Mitsuhashi, Ethan Firestone, Aimee~F Luat, Neena~I Marupudi, Sandeep Sood, and Eishi Asano.
\newblock Developmental atlas of phase-amplitude coupling between physiologic high-frequency oscillations and slow waves.
\newblock {\em Nature Communications}, 14(1):6435, 2023.

\bibitem{he2022masked}
Kaiming He, Xinlei Chen, Saining Xie, Yanghao Li, Piotr Doll{\'a}r, and Ross Girshick.
\newblock Masked autoencoders are scalable vision learners.
\newblock In {\em Proceedings of the IEEE/CVF conference on computer vision and pattern recognition}, pages 16000--16009, 2022.

\bibitem{li2021neural}
Adam Li, Chester Huynh, Zachary Fitzgerald, Iahn Cajigas, Damian Brusko, Jonathan Jagid, Angel~O Claudio, Andres~M Kanner, Jennifer Hopp, Stephanie Chen, et~al.
\newblock Neural fragility as an {EEG} marker of the seizure onset zone.
\newblock {\em Nature Neuroscience}, 24(10):1465--1474, 2021.

\bibitem{fedele2017resection}
Tommaso Fedele, Sergey Burnos, Ece Boran, Niklaus Krayenb{\"u}hl, Peter Hilfiker, Thomas Grunwald, and Johannes Sarnthein.
\newblock Resection of high frequency oscillations predicts seizure outcome in the individual patient.
\newblock {\em Scientific Reports}, 7(1):13836, 2017.

\bibitem{ds003555}
Dorottya Cserpan, Ece Boran, Richard Rosch, San Pietro~Lo Biundo, Georgia Ramantani, and Johannes Sarnthein.
\newblock Dataset of {EEG} recordings of pediatric patients with epilepsy based on the 10-20 system, 2021.

\bibitem{ds003844}
Matteo Demuru, Dorien van Blooijs, Willemiek Zweiphenning, Dora Hermes, Frans Leijten, Maeike Zijlmans, and RESPect group.
\newblock A practical workflow for organizing clinical intraoperative and long-term {iEEG} data in {BIDS}.
\newblock {\em Neuroinformatics}, 20(3):727--736, 2022.

\bibitem{ds003876}
Kristin Gunnarsdottir, Adam Li, Rachel Smith, Joon Kang, Anna Korzeniewska, Nathan Crone, Adam Rouse, Jennifer Cheng, Michael Kinsman, Patrick Landazuri, Utku Uysal, Carol Ulloa, Nathaniel Cameron, Iahn Cajigas, Jonathan Jagid, Andres Kanner, Turki Elarjani, Manuel Bicchi, Sara Inati, Kareem Zaghloul, Varina Boerwinkle, Sarah Wyckoff, Niravkumar Barot, Jorge Gonzalez-Martinez, and Sridevi Sarma.
\newblock Epilepsy {iEEG} interictal multicenterdataset, 2023.

\bibitem{ds004100}
John~M Bernabei, Nishant Sinha, T~Campbell Arnold, Erin Conrad, Ian Ong, Akash~R Pattnaik, Joel~M Stein, Russell~T Shinohara, Timothy~H Lucas, Dani~S Bassett, et~al.
\newblock Normative intracranial {EEG} maps epileptogenic tissues in focal epilepsy.
\newblock {\em Brain}, 145(6):1949--1961, 2022.

\bibitem{dimakopoulos2022information}
Vasileios Dimakopoulos, Pierre M{\'e}gevand, Lennart~H Stieglitz, Lukas Imbach, and Johannes Sarnthein.
\newblock Information flows from hippocampus to auditory cortex during replay of verbal working memory items.
\newblock {\em eLife}, 11:e78677, 2022.

\bibitem{zhang2022refining}
Yipeng Zhang, Qiujing Lu, Tonmoy Monsoor, Shaun~A Hussain, Joe~X Qiao, Noriko Salamon, Aria Fallah, Myung~Shin Sim, Eishi Asano, Raman Sankar, et~al.
\newblock Refining epileptogenic high-frequency oscillations using deep learning: A reverse engineering approach.
\newblock {\em Brain Communications}, 4(1):fcab267, 2022.

\bibitem{veloso2017big}
L~Veloso, J~McHugh, E~Von~Weltin, S~Lopez, I~Obeid, and J~Picone.
\newblock Big data resources for {EEGs}: Enabling deep learning research.
\newblock In {\em 2017 IEEE Signal Processing in Medicine and Biology Symposium (SPMB)}, pages 1--3. IEEE, 2017.

\bibitem{shah2018temple}
Vinit Shah, Eva Von~Weltin, Silvia Lopez, James~Riley McHugh, Lillian Veloso, Meysam Golmohammadi, Iyad Obeid, and Joseph Picone.
\newblock The temple university hospital seizure detection corpus.
\newblock {\em Frontiers in Neuroinformatics}, 12:83, 2018.

\bibitem{detti2020eeg}
Paolo Detti, Giampaolo Vatti, and Garazi Zabalo Manrique~de Lara.
\newblock {EEG} synchronization analysis for seizure prediction: A study on data of noninvasive recordings.
\newblock {\em Processes}, 8(7):846, 2020.

\bibitem{nasreddine2021epileptic}
Wassim Nasreddine.
\newblock Epileptic {EEG} dataset.
\newblock {\em Mendeley Data}, 1, 2021.

\bibitem{falach2024annotated}
Rotem Falach, Maya Geva-Sagiv, Dawn Eliashiv, Lilach Goldstein, Ofer Budin, Guy Gurevitch, Genela Morris, Ido Strauss, Amir Globerson, Firas Fahoum, et~al.
\newblock Annotated interictal discharges in intracranial {EEG} sleep data and related machine learning detection scheme.
\newblock {\em Scientific Data}, 11(1):1354, 2024.

\bibitem{shoeb2009application}
Ali~Hossam Shoeb.
\newblock {\em Application of machine learning to epileptic seizure onset detection and treatment}.
\newblock PhD thesis, Massachusetts Institute of Technology, 2009.

\bibitem{fasil2021scalp}
OK~Fasil, R~Rajesh, and Rajith~K Ravindren.
\newblock Scalp {EEG} recordings of pediatric epilepsy patients: A dataset for automatic detection of interictal epileptiform discharges from routine {EEG}.
\newblock {\em Data in Brief}, 39:107680, 2021.

\bibitem{stevenson2019dataset}
Nathan~J Stevenson, Karoliina Tapani, Leena Lauronen, and Sampsa Vanhatalo.
\newblock A dataset of neonatal {EEG} recordings with seizure annotations.
\newblock {\em Scientific Data}, 6(1):1--8, 2019.

\bibitem{nejedly2020multicenter}
Petr Nejedly, Vaclav Kremen, Vladimir Sladky, Jan Cimbalnik, Petr Klimes, Filip Plesinger, Filip Mivalt, Vojtech Travnicek, Ivo Viscor, Martin Pail, et~al.
\newblock Multicenter intracranial {EEG} dataset for classification of graphoelements and artifactual signals.
\newblock {\em Scientific Data}, 7(1):179, 2020.

\bibitem{collins2024tripod+}
Gary~S Collins, Karel~GM Moons, Paula Dhiman, Richard~D Riley, Andrew~L Beam, Ben Van~Calster, Marzyeh Ghassemi, Xiaoxuan Liu, Johannes~B Reitsma, Maarten Van~Smeden, et~al.
\newblock Tripod+ ai statement: updated guidance for reporting clinical prediction models that use regression or machine learning methods.
\newblock {\em bmj}, 385, 2024.

\bibitem{su2024roformer}
Jianlin Su, Murtadha Ahmed, Yu~Lu, Shengfeng Pan, Wen Bo, and Yunfeng Liu.
\newblock Roformer: Enhanced transformer with rotary position embedding.
\newblock {\em Neurocomputing}, 568:127063, 2024.

\bibitem{kornblith2019similarity}
Simon Kornblith, Mohammad Norouzi, Honglak Lee, and Geoffrey Hinton.
\newblock Similarity of neural network representations revisited.
\newblock In {\em International conference on machine learning}, pages 3519--3529. PMLR, 2019.

\bibitem{goswami2024moment}
Mononito Goswami, Konrad Szafer, Arjun Choudhry, Yifu Cai, Shuo Li, and Artur Dubrawski.
\newblock {MOMENT}: A family of open time-series foundation models.
\newblock {\em arXiv preprint arXiv:2402.03885}, 2024.

\bibitem{nie2022time}
Yuqi Nie, Nam~H Nguyen, Phanwadee Sinthong, and Jayant Kalagnanam.
\newblock A time series is worth 64 words: Long-term forecasting with transformers.
\newblock {\em arXiv preprint arXiv:2211.14730}, 2022.

\bibitem{tangermann2012review}
Michael Tangermann, Klaus-Robert M{\"u}ller, Ad~Aertsen, Niels Birbaumer, Christoph Braun, Clemens Brunner, Robert Leeb, Carsten Mehring, Kai~J Miller, Gernot~R M{\"u}ller-Putz, et~al.
\newblock Review of the bci competition iv.
\newblock {\em Frontiers in neuroscience}, 6:55, 2012.

\bibitem{goldberger2000physiobank}
Ary~L Goldberger, Luis~AN Amaral, Leon Glass, Jeffrey~M Hausdorff, Plamen~Ch Ivanov, Roger~G Mark, Joseph~E Mietus, George~B Moody, Chung-Kang Peng, and H~Eugene Stanley.
\newblock Physiobank, physiotoolkit, and physionet: components of a new research resource for complex physiologic signals.
\newblock {\em circulation}, 101(23):e215--e220, 2000.

\bibitem{song2022eeg}
Yonghao Song, Qingqing Zheng, Bingchuan Liu, and Xiaorong Gao.
\newblock {EEG Conformer}: Convolutional transformer for {EEG} decoding and visualization.
\newblock {\em IEEE Transactions on Neural Systems and Rehabilitation Engineering}, 31:710--719, 2022.

\bibitem{jiang2024csp}
Xue Jiang, Lubin Meng, Xinru Chen, Yifan Xu, and Dongrui Wu.
\newblock {CSP-Net}: Common spatial pattern empowered neural networks for {EEG}-based motor imagery classification.
\newblock {\em Knowledge-Based Systems}, 305:112668, 2024.

\bibitem{santamaria2020eeg}
Eduardo Santamaria-Vazquez, Victor Martinez-Cagigal, Fernando Vaquerizo-Villar, and Roberto Hornero.
\newblock {EEG-Inception}: A novel deep convolutional neural network for assistive {ERP}-based brain-computer interfaces.
\newblock {\em IEEE Transactions on Neural Systems and Rehabilitation Engineering}, 28(12):2773--2782, 2020.

\bibitem{lawhern2018eegnet}
Vernon~J Lawhern, Amelia~J Solon, Nicholas~R Waytowich, Stephen~M Gordon, Chou~P Hung, and Brent~J Lance.
\newblock {EEGNet}: A compact convolutional neural network for {EEG}-based brain--computer interfaces.
\newblock {\em Journal of Neural Engineering}, 15(5):056013, 2018.

\bibitem{supratak2020tinysleepnet}
Akara Supratak and Yike Guo.
\newblock {TinySleepNet}: An efficient deep learning model for sleep stage scoring based on raw single-channel {EEG}.
\newblock In {\em 2020 42nd Annual International Conference of the IEEE Engineering in Medicine \& Biology Society (EMBC)}, pages 641--644. IEEE, 2020.

\bibitem{altaheri2022physics}
Hamdi Altaheri, Ghulam Muhammad, and Mansour Alsulaiman.
\newblock Physics-informed attention temporal convolutional network for {EEG}-based motor imagery classification.
\newblock {\em IEEE Transactions on Industrial Informatics}, 19(2):2249--2258, 2022.

\bibitem{chen2025df}
Guibin Chen, Gang Li, Wanxiu Xu, Hanfan Wu, Suhong Ye, and Bin Zhou.
\newblock A {DF-SSA} analytical framework for revealing variations in multidimensional {EEG} features of epileptic seizures.
\newblock {\em Biomedical Signal Processing and Control}, 100:107073, 2025.

\end{thebibliography}

\clearpage

\clearpage

\begin{table*}[h]
\centering
\small
\caption{Summary of pre-training datasets}
\label{tab:pretrain ds}
\resizebox{\textwidth}{!}{%
\begin{tabular}{cccccccc}
\toprule
\textbf{Index} & \textbf{Dataset Name}  & \textbf{Recording Durations (hours)} & \textbf{Sampling Rates (Hz)} & \textbf{Signal Types} & \textbf{References} \\ 
\midrule
1 & DS003029            & 8.2      & 1024, 1000, 2000, 500, 250   & SEEG, ECoG  & \cite{li2021neural} \\ 
2 & DS003498            & 32.3     & 2000                         & ECoG        & \cite{fedele2017resection} \\ 
3 & DS003555            & 89.0     & 1024, 2048                   & EEG         & \cite{ds003555} \\ 
4 & DS003844            & 2.7      & 256, 2048                    & ECoG        & \cite{ds003844} \\ 
5 & DS003876            & 5.7      & 1024, 512, 1000, 2000, 500   & SEEG, ECoG  & \cite{ds003876} \\ 
6 & DS004100            & 25.7     & 512, 1024, 500, 256          & ECoG, SEEG  & \cite{ds004100} \\ 
7 & DS004752            & 0.3      & 200, 4096, 4000, 2000        & ECoG, EEG, SEEG & \cite{dimakopoulos2022information} \\ 
8 & DS005398           & 90.9      & 1000, 200, 2000             & SEEG, ECoG     & \cite{zhang2022refining} \\ 
9 & TUEP               & 530.3      & 256, 512, 1000, 400, 250   & EEG    & \cite{veloso2017big}     \\ 
10 & TUSZ               & 1473.5      & 256, 512, 1000, 400, 250  & EEG    & \cite{shah2018temple} \\ 
11 & Siena              & 128.4      & 512      & EEG    & \cite{detti2020eeg} \\ 
12 & ZJNU          & 269.3      & 500      & EEG    & \cite{chen2025df} \\ 
13 & AUB            & 57.8      & 500      & EEG    & \cite{nasreddine2021epileptic} \\ 
\midrule
   & Summarize     & 2741.1  & 200 - 4096 Hz & EEG, SEEG, ECoG & \\
\bottomrule

\end{tabular}
}
\end{table*}

\begin{table*}[h]
\centering
\caption{Summary of downstream tasks}
\label{tab:downstream ds}
\resizebox{\textwidth}{!}{%
\begin{tabular}{ccccc}
\toprule
\textbf{Dataset Name} & \textbf{Task} & \textbf{Signal Types} & \textbf{Classification Task} & \textbf{References} \\ 
\midrule
TASMC-UCLA & SOZ Localization       & SEEG      & SOZ activity v.s. Non-SOZ activity     & \cite{falach2024annotated} \\ 
CHB-MIT & Seizure Prediction          & EEG      & Inter-Ictal activity v.s. Pre-Ictal activity   & \cite{shoeb2009application}  \\ 
CUK-IMHANS & IED Detection          & EEG      & Activity Including IED  v.s. Activity Not Including IED    & \cite{fasil2021scalp} \\ 
FNUSA & Pathology Detection      & SEEG      & Pathology Activity v.s. Physiological Activity     & \cite{nejedly2020multicenter} \\ 
MAYO & Pathology Detection       & SEEG      & Pathology Activity v.s. Physiological Activity     & \cite{nejedly2020multicenter}  \\
HUH & Seizure Detection         & EEG      & Ictal Activity v.s. Inter-Ictal Activity   & \cite{stevenson2019dataset}  \\ 
\bottomrule
\end{tabular}
}
\end{table*}

\begin{table*}[h]
    \centering
    \caption{F1 score comparisons of EpiNT against pre-trained models}
    \label{tab: compare, pre-trained}
    \resizebox{\textwidth}{!}{%
    \begin{tabular}{p{2.2cm}lllllll}
    \toprule
    \textbf{Fine-tuning Strategy} & \textbf{Model Name} & \textbf{TASMC-UCLA} & \textbf{CHB-MIT} & \textbf{HUH} & \textbf{CUK-IMHANS} & \textbf{FNUSA} & \textbf{MAYO} \\ 
    \midrule
    \multirow{6}{*}{Linear-Probing}
    &MAE-EEG         & $0.875\pm0.039$               & $0.820\pm0.035$           & $0.739\pm0.005$***        & $0.719\pm0.088$*          & $0.734\pm0.028$*                       & $0.839\pm0.086$  \\
    &Brant           & $0.876\pm0.037$               & $0.740\pm0.007$           & $0.715\pm0.018$**         & $0.728\pm0.023$**         & $0.720\pm0.026$***                       & $0.872\pm0.051$           \\
    &MOMENT          & $0.896\pm0.035$               & $0.779\pm0.051$           & $0.728\pm0.008$***        & $0.765\pm0.069$*          & $0.752\pm0.024$                       & $0.907\pm0.052$            \\
    &PatchTST        & $0.887\pm0.040$               & $0.765\pm0.022$*          & $0.733\pm0.001$****       & $0.843\pm0.016$*          & $0.669\pm0.094$                       & $0.835\pm0.099$          \\
    &VQ-MTM          & $0.908\pm0.010$               & $0.836\pm0.012$           & $0.767\pm0.003$*          & $0.849\pm0.034$           & $0.735\pm0.066$                       & $0.929\pm0.018$       \\
    &EpiNT            & $\mathbf{0.917\pm0.014}$      & $\mathbf{0.857\pm0.056}$  & $\mathbf{0.770\pm0.003}$  & $\mathbf{0.851\pm0.035}$  & $\mathbf{0.805\pm0.038}$              & $\mathbf{0.930\pm0.027}$         \\
    \midrule
    \multirow{6}{*}{Last-Layer}
    &MAE-EEG                     & $0.898\pm0.019$*              & $0.868\pm0.007$*                  & $0.844\pm0.004$                   & $0.813\pm0.015$                   & $0.757\pm0.008$***                    & $0.942\pm0.003$***    \\
    &Brant                       & $0.915\pm0.014$               & $0.868\pm0.013$                   & $\mathbf{0.848\pm0.014}$          & $0.817\pm0.019$                   & $0.816\pm0.012$***                    & $0.932\pm0.012$**          \\
    &MOMENT                      & $0.912\pm0.014$               & $0.866\pm0.023$                   & $0.847\pm0.012$                   & $0.811\pm0.011$*                  & $0.842\pm0.006$***                    & $0.925\pm0.012$**            \\
    &PatchTST                    & $0.905\pm0.017$               & $\mathbf{0.882\pm0.015}$          & $0.820\pm0.016$*                  & $0.804\pm0.015$                   & $0.827\pm0.004$****                   & $0.921\pm0.016$***          \\
    &VQ-MTM                      & $0.912\pm0.008$*              & $0.857\pm0.011$                   & $0.843\pm0.010$                   & $0.807\pm0.038$                   & $0.863\pm0.005$***                    & $0.940\pm0.003$       \\
    &EpiNT                        & $\mathbf{0.924\pm0.006}$      & $0.862\pm0.010$                   & $0.842\pm0.008$                   & $\mathbf{0.846\pm0.027}$          & $\mathbf{0.885\pm0.005}$              & $\mathbf{0.959\pm0.003}$        \\
    \midrule
    \multirow{6}{*}{Full-Parameter}
    &MAE-EEG                     & $0.819\pm0.061$*              & $0.911\pm0.002$               & $\mathbf{0.866\pm0.003}$*     & $0.841\pm0.022$                               & $0.788\pm0.009$****                   & $0.953\pm0.002$  \\
    &Brant                       & $0.921\pm0.017$               & $0.900\pm0.008$*              & $0.846\pm0.018$               & $0.831\pm0.025$                               & $0.872\pm0.005$**                     & $0.960\pm0.003$**          \\
    &MOMENT                      & $0.883\pm0.026$*              & $0.902\pm0.005$**             & $0.858\pm0.011$               & $\mathbf{0.846\pm0.020}$                      & $0.876\pm0.002$*                      & $0.954\pm0.007$*           \\
    &PatchTST                    & $0.908\pm0.011$*              & $0.896\pm0.002$**             & $0.845\pm0.003$**             & $0.817\pm0.015$                               & $0.873\pm0.006$*                      & $0.955\pm0.008$*          \\
    &VQ-MTM                      & $0.921\pm0.014$               & $0.900\pm0.003$****           & $0.855\pm0.006$               & $0.828\pm0.021$                               & $0.877\pm0.002$*                      & $0.957\pm0.003$**      \\
    &EpiNT                        & $\mathbf{0.925\pm0.009}$      & $\mathbf{0.913\pm0.003}$      & $0.855\pm0.004$               & $0.811\pm0.032$                               & $\mathbf{0.886\pm0.005}$              & $\mathbf{0.965\pm0.003}$        \\
    \bottomrule
    \end{tabular}
    }
    \parbox{\textwidth}{\footnotesize Data are presented as mean $\pm$ std. Statistical significance: *: p-value$<$0.05; **: p-value$<$0.01; ***: p-value$<$0.001; ****: p-value$<$0.0001.}
\end{table*}

\end{document}